\documentclass[prd,aps,a4paper,twocolumn,footinbib,superscriptaddress,showpacs]{revtex4}
\usepackage{graphicx}
\usepackage{color}
\usepackage{nicefrac}
\usepackage{dcolumn}
\usepackage{bm}
\usepackage{slashed}
\usepackage{latexsym}
\usepackage{textcomp,amsmath,amssymb}

\begin{document}

\title{Improved fast-rotating black hole evolution simulations with modified Baumgarte-Shapiro-Shibata-Nakamura formulation}
\author{Hwei-Jang Yo}
\affiliation{Department of Physics, National Cheng-Kung University,
Tainan 701, Taiwan}
\author{Zhoujian Cao}
\affiliation{Institute of Applied Mathematics and LSEC, Academy of
Mathematics and Systems Science, Chinese Academy of Sciences,
Beijing 100190, China}
\author{Chun-Yu Lin}
\affiliation{National Center for High-Performance Computing,
Hsinchu 300, Taiwan}
\author{Hsing-Po Pan}
\affiliation{Department of Physics, National Cheng-Kung University,
Tainan 701, Taiwan}

\date{\today}

\begin{abstract}
Different formulations of Einstein's equations used in numerical relativity can
affect not only the stability but also the accuracy of numerical simulations.
In the original Baumgarte-Shapiro-Shibata-Nakamura (BSSN) formulation,
the loss of the angular momentum, $J$, is
non-negligible in highly spinning single black hole evolutions.
This loss also appears, usually right after the merger, in highly spinning
binary black hole simulations,
The loss of $J$ may be attributed to some unclear numerical dissipation.
Reducing unphysical dissipation is expected to result in more stable and
accurate evolutions.
In the previous work \cite{yhlc12} we proposed several modifications which
are able to prevent black hole evolutions from the unphysical dissipation,
and the resulting simulations are more stable than in the traditional BSSN
formulation.
Specifically, these three modifications (M1, M2, and M3) enhance the effects
of stability, hyperbolicity, and dissipation of the formulation.
We experiment further in this work with these modifications, and demonstrate
that these modifications improve the accuracy and also effectively suppress the
loss of $J$, particularly in the black hole simulations with the initially
large ratio of $J$ and the square of the ADM mass.
\end{abstract}

\maketitle
\section{Introduction}
Development of numerical relativity has been rapid
after the breakthroughs in 2005 and 2006 (see, e.g., \cite{pref05,NR06a,NR06b}).
Numerical relativity has now become an indispensable and effective tool
in the research of general relativity and relativistic astrophysics.
It has been extensively studied in several areas; and applied to the construction of
gravitational waveform template banks for detection \cite{PanBuoBuc10},
to the kick phenomena of general binary systems
\cite{BHkicksa,BHkicksb,BHkicksc,BHkicksd,BHkickse}, and to astrophysical
problems such as the equation of state of neutron stars \cite{ReaMarShi09,PhysRevD.85.104030,PhysRevD.86.044030,PhysRevD.87.044001},
electromagnetic counterparts of gravitational waves \cite{JetBHa,JetBHb},
gamma ray bursts \cite{KiuSekShi10}, accretion disks \cite{SekShi11} and so on.
These applications and numerical investigations demand increasingly greater accuracy, besides stability;
thus it is important and necessary to unremittingly refine the formulations and schemes.

Among the methods to enhance the stability and accuracy in numerical
relativity, the $3+1$ formulation of Einstein's equations is favored by many researchers.
The Baumgarte--Shapiro--Shibata--Nakamura (BSSN) formulation
\cite{BSSN95_99a,BSSN95_99b} is the most popular scheme, and it is usually
implemented in first-order-in-time and second-order-in-space
finite-differencing codes.
Many works have been focused on modifying the original BSSN formulation to
achieve better numerical stability and accuracy
\cite{almb01,yhbs02,ygsh02,kksh08,ttys12,bjet12}.
For example, borrowing from ideas in the Generalized Harmonic (GH)
formulation \cite{pref05,lial06}, the Z4 conformal (Z4c) formulation \cite{Z4ca} and the traceless-conformal and covariant Z4
(CCZ4) formulation \cite{CCZ4a}
both show good constraint damping behavior \cite{CCZ4b,Z4cb,Z4cc}.

There is still room to improve the BSSN formulation to obtain better stability and accuracy, e.g., see references
\cite{almb01,yhbs02,LaPS02,ygsh02,czyy08,etienne2014improved}.
In \cite{yhlc12}, we adopt a different approach from the CCZ4 and Z4c
formulations to modify the BSSN formulation.
The basic idea is to suppress the numerical error by adding combinations of
constraint terms to the field equations in the BSSN formulation
without changing the solution analytically, to modify the leading terms of
the field equations.
And we demonstrated that our modifications achieved more stable simulations
than the traditional BSSN formulation.
Specifically, our last work \cite{yhlc12} has shown improvements in
constraint damping and in the late-time behavior of the gravitational waveforms.
In this work, we would like to emphasize the effectiveness of these
modifications on the evolution of the black holes with higher spins,
hoping to meet the demand of modeling extreme sources for
the gravitational wave detection.

\begin{figure}[thbp]
\begin{tabular}{c}
\includegraphics[width=0.98\columnwidth]{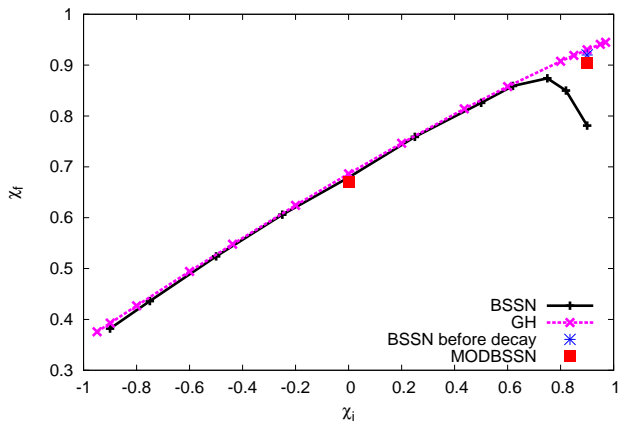}
\end{tabular}
\caption{
Dimensionless spin parameter $\chi_f$ of the final black hole as a function
of the initial dimensionless spin parameter $\chi_i$ of the individual black
hole in the binary black hole evolution for the BSSN and GH formulations.
The data 'BSSN' given in \cite{MarTicBru08} are calculated through the BAM code with BSSN formulation and the finite-differencing
method. The data 'GH' given in \cite{HemLovLor13} are calculated through the SpEC code with GH formulation and the spectral method.
The data `BSSN before decay' given in \cite{MarTicBru08} is calculated through the LEAN code \cite{MarTicBru08}.
It corresponds to the result right after the final black hole forms,
and the data `MODBSSN' are calculated through AMSS-NCKU code in current work, and the modified BSSN formulation (check the main text for more detail) is used.}
\label{fig1}
\end{figure}

It was found in \cite{MarTicBru08} that the
angular momentum decays right after the final black hole forms in the binary
black hole evolution simulations.
For the single spinning black hole, the angular momentum also decays
when the dimensionless spin $s/m^2>0.75$ (compare the cases
$s/m^2=0.53$ and $s/m^2=0.9$ in Fig.~4 of \cite{MarTicBru08}).
This decay is neither due to the resolution nor the initial separation of
the binary \cite{MarTicBru08}.
In contrast, the result from the SpEC code does not show this tendency
\cite{HemLovLor13}.
For comparing those results, we plot in Fig.~\ref{fig1}
$\chi_f\equiv s_f/m_f^2$ of the final black hole as a function of the
initial spin parameter $\chi_i\equiv s_i/m_i^2$ for individual black hole
component.
We find that they are consistent when $\chi_i<0.75$ and, when $\chi_i$
getting larger, $\chi_f$ in the traditional BSSN formulation becomes smaller
than that in the GH formulation.
The difference should not be attributed to whether the spectral method or the finite-differencing
method is used.
However, it is yet unclear if the issue comes from the formulation
itself or from the puncture method.
In this work, we therefore try to resolve this problem by simulating single
and binary black hole evolutions with the modified BSSN formulation as
proposed in \cite{yhlc12}.
We will show that the angular momentum is more
accurate and its conservation is much better than in the traditional BSSN formulation.
Via the better conservation of the angular momentum, it is also expected that
the accuracy of the other related physical quantities will be improved at the
same time with the modified formulation.

The rest of this work is organized as follows:
In the next section, we will give an explicit description of the modifications
to the BSSN formulation, discuss the related accuracy problems, and describe
the numerical implementation.
We then report on the test results on single spinning black hole in
Sec.~\ref{sec::IIIA}.
The results for highly spinning binary black hole are presented in
Sec.~\ref{sec::IIIB}.
And the discussion and summary will be presented in the Sec.~\ref{sec::IV}.
Throughout the paper, geometric units with $G=c=1$ are used.
Einstein summation rule is adopted unless stated explicitly.
\section{Modifications and Numerical Implementations}\label{sec::II}
The BSSN formulation and the numerical recipes for implementation have been
described in details in previous articles \cite{yhbs02,czyy08}.
Here we only mention several major steps that have usually been adopted
\cite{almb01,LaPS02} in the {\it traditional} BSSN formulation:
\begin{itemize}
\item
In order to enforce the algebraic constraints of the unimodular determinant
 of $\tilde\gamma_{ij}$, i.e., $\tilde\gamma=1$, and of the tracelessness of
$\tilde A_{ij}$,
i.e., $\tilde{\gamma}^{ij}\tilde{A}_{ij} = 0$,
the numerical values of
$\tilde\gamma_{ij}$ and $\tilde A_{ij}$ are replaced with
$\tilde\gamma_{ij}\rightarrow\tilde\gamma^{-1/3}\tilde\gamma_{ij}$,
$\tilde A_{ij}\rightarrow\tilde A_{\langle ij\rangle}$ after every time step,
wherein the two indices in the angle bracket $\langle\rangle$ is taken to be its
symmetric and traceless part.
\item
The conformal connection functions $\tilde\Gamma^i$ are promoted to be
independent variables in the BSSN formulation, which leads to the
$\Gamma$-constraints
${\mathcal G}^i \equiv \tilde{\Gamma}^i-\tilde{\Gamma}^i_{\bf g}=0$,
where $\tilde{\Gamma}^i_{\bf g}\equiv\tilde{\gamma}^{jk}
\tilde{\Gamma}^i{}_{jk}$.
The conventional approach to enforce the $\Gamma$-constraints is to replace
all the undifferentiated $\tilde\Gamma^i$ with $\tilde\Gamma^i_{\bf g}$.
\item
The high-order Kreiss-Oliger (KO) method is employed to dissipate effectively
the numerical noise.
\end{itemize}
These traditional approaches with suitable gauge condition have enabled
fruitful studies on the black hole problem.
Yet earlier investigations, e.g., \cite{MarTicBru08}, indicated that,
in some near extreme situation, the traditional BSSN formulation is not
robust enough to conserve the constraints and global quantities.
We plan to test the following modifications which have been introduced in
\cite{yhlc12}, and compare the results from our modifications
with those from the traditional BSSN formulation.
The three proposed modifications are as follows:
\subsection{Modification M1}
Instead of replacing all the undifferentiated $\tilde\Gamma^i$ with
$\tilde\Gamma^i_{\bf g}$,
M1 modifies the conformal 3-connection appearing in the right-hand-side of all
the field equations, and changes the linear terms in the field equation of
${\tilde\Gamma}^i$.
The new conformal 3-connection in all field equations now takes the form
\begin{equation}
{\tilde\Gamma}^i{}_{jk}\rightarrow{\tilde\Gamma}^i{}_{jk}
-\frac{3}{5}\delta^i{}_{\langle j}{\tilde T}_{k\rangle}
-\frac{1}{5}\delta^i{}_{\langle j}{\mathcal G}_{k\rangle}+\frac{1}{3}
{\tilde\gamma}_{jk}{\mathcal G}^i,\label{newG}
\end{equation}
where ${\tilde T}_i\equiv{\tilde\Gamma}^k{}_{ki}=(\ln\sqrt{\tilde\gamma})_{,i}$
vanishes analytically, but could be nonzero due to numerical error.
This expression is motivated by the unique algebraic decomposition for any
third-rank symmetric tensor with two indices.
See \cite{yhlc12} for the details.

To change the behavior of the linear term in the field equation of
${\tilde\Gamma}^i$,
we replace the original field equations of $\tilde{\Gamma}^i$ with
\begin{align}
\partial_t\tilde{\Gamma}^i=&2\alpha[{\tilde\Gamma}^i{}_{jk}\tilde A^{jk}-
\frac{2}{3}({\tilde\gamma}^{ij}K){}_{,j}+6\tilde{A}^{ij}\phi_{,j}]-
2\tilde{A}^{ij}\alpha_{,j}\nonumber\\
 &+\beta^j\tilde{\Gamma}^i{}_{,j}-\tilde{\Gamma}^j\beta^i{}_{,j}
  +\tilde{\gamma}^{jk}\beta^i{}_{,jk}+\frac{1}{3}\tilde{\gamma}^{ij}
 \beta^k{}_{,jk} \nonumber\\
  &+\frac{2}{3}(\beta^k{}_{,k}-2\alpha K)\tilde{\Gamma}^i
  -(1+\xi)\Theta(\lambda^i)\lambda^i{\mathcal G}^i,\label{dtG}
\end{align}
wherein $\Theta(x)$ is the step function
\begin{equation}
\Theta(x)=\left\{\begin{array}{cc}0&\text{ if } x < 0\\
1&\text{ if } x > 0\end{array}\right.
\end{equation}
and $\lambda^i$ is
\begin{equation}
\lambda^i=\frac{2}{3}(\beta^k{}_{,k}-2\alpha K)-\beta^{\hat i}{}_{,\hat i}
-\frac{2}{5}\alpha{\tilde A}_{\hat i}{}^{\hat i}.
\end{equation}
Note that the index with hat, i.e., $\hat i$, means that no index summation
is carried out with respect to this index.
$\xi$ is chosen to be $1$ in all cases in this work.
This modification plays an indispensable role in the whole
modification scheme to enhance both the stability and accuracy of the system.

Notice that there is one term in eqn (\ref{dtG}) including a step function,
i.e., $(1+\xi)\Theta(\lambda^i)\lambda^i{\mathcal G}^i$.
Due to its switch character and the possible sign fluctuation of the numerical
value of its argument $\lambda^i$ when $\lambda^i$ is close to zero,
the step function should be sensitive to the resolution used in simulations.
So we expect that this modification could affect majorly the numerical
convergence behavior of the modified BSSN formulation.

\subsection{Modification M2}
The idea behind M2 is similar to that in obtaining Eq.~(\ref{newG}).
The algebraic structure of $\partial_t\tilde{\gamma}_{ij}$, similar to the
algebraic structure of the conformal 3-connection, allows us to write
the ${\tilde\gamma}_{ij}$-field equation as
\begin{equation}\label{dtg2}
\partial_t\tilde{\gamma}_{ij}\rightarrow\partial_t\tilde{\gamma}_{ij}
+\sigma\beta_{(i}{\mathcal G}_{j)}
-\frac{1}{5}{\tilde\gamma}_{ij}\beta^k{\mathcal G}_k;
\end{equation}
and we set $\sigma=1/10$ in this work.
This modification enhances the hyperbolicity of the system and propagates the
constraint violation residual away effectively.
\subsection{Modification M3}
This dissipation type of modification M3 is motivated from \cite{ygsh02}.
The major difference is that we use the symmetric traceless part of the partial derivative of the
momentum constraint (instead of the symmetric part of
its covariant derivative as in \cite{ygsh02}) to re-write the
${\tilde A}_{ij}$-field equation as
\begin{equation}\label{modta}
\partial_t\tilde{A}_{ij}\rightarrow\partial_t\tilde{A}_{ij}+h^2
{\mathcal M}_{\langle i,j\rangle},
\end{equation}
wherein ${\mathcal M}_i$ is the momentum constraint, $h$ is the grid width.
This modification provides a dissipation mechanism on ${\tilde A}_{ij}$,
and serves as a natural alternative to the KO dissipation.
In this work, we apply this modification instead of the KO method to check
its capability in dissipation and also compare its effect with KO's.

There is a concern about the convergence of the whole system with this
modification.
At first glance, Equation (\ref{modta}) might change the convergence order
of a system to be only second-order accurate at most since the addtion term in
the equation, i.e., $h^2{\mathcal M}_{\langle i,j\rangle}$, is only
proportional explicitly to $h^2$.
However, this is not the case.
If one system is $p^{\rm th}$-order convergent,
then the momentum constraint ${\mathcal M}_i\simeq 0$ will converge to zero
with the rate of $h^p$. So will the term ${\mathcal M}_{\langle i,j\rangle}$.
Therefore, if we combine the convergence order of
${\mathcal M}_{\langle i,j\rangle}$ and the the multiplier $h^2$,
the term introduced in eqn (\ref{modta}) will converge to zero with the rate
$h^{p+2}$, which convergence is faster than the rest of the system.
Thus this modification will not reduce the convergence order of the system
{\it analytically}.

\subsection{Numerical Implementation}
The AMSS-NCKU code with the standard moving box style mesh refinement
\cite{czyy08,cao10,yhlc12} is used in this work.
We used 10 mesh levels, all of which are fixed in the cases of
single black hole evolution, and the finest 3 levels are
movable in evolving the binary black holes (BBHs).
In each fixed level, we used one box with $128\times 128\times 64$ grids
with assumed equatorial symmetry.
The outermost physical boundary is $512 M$ and this makes the finest resolution to
be $h=M/64$.
For the movable levels, two boxes with $64\times 64\times 32$ grids are used
to cover each black hole.
In time direction, the Berger-Oliger numerical scheme is adopted for the
levels higher than four.

The moving puncture gauge condition
\begin{eqnarray}
&&\partial_t\alpha=\beta^i\alpha_{,i}-2\alpha K,\\
&&\partial_t\beta^i=\frac{3}{4}B^i+\beta^j\beta^i_{,j},\\
&&\partial_tB^i=\partial_t\tilde{\Gamma}^i-\eta B^i+\beta^jB^i_{,j}-
\beta^j\tilde{\Gamma}^i_{,j}.
\end{eqnarray}
is used and has been shown to give good behavior for the black hole
simulations in \cite{czyy08}.
In this paper we use $\eta=2 M$ with $M$ being the ADM mass of the given
configuration.
\section{Numerical results}\label{sec::III}
\begin{figure*}[thbp]
\begin{tabular}{rl}
\includegraphics[width=\columnwidth]{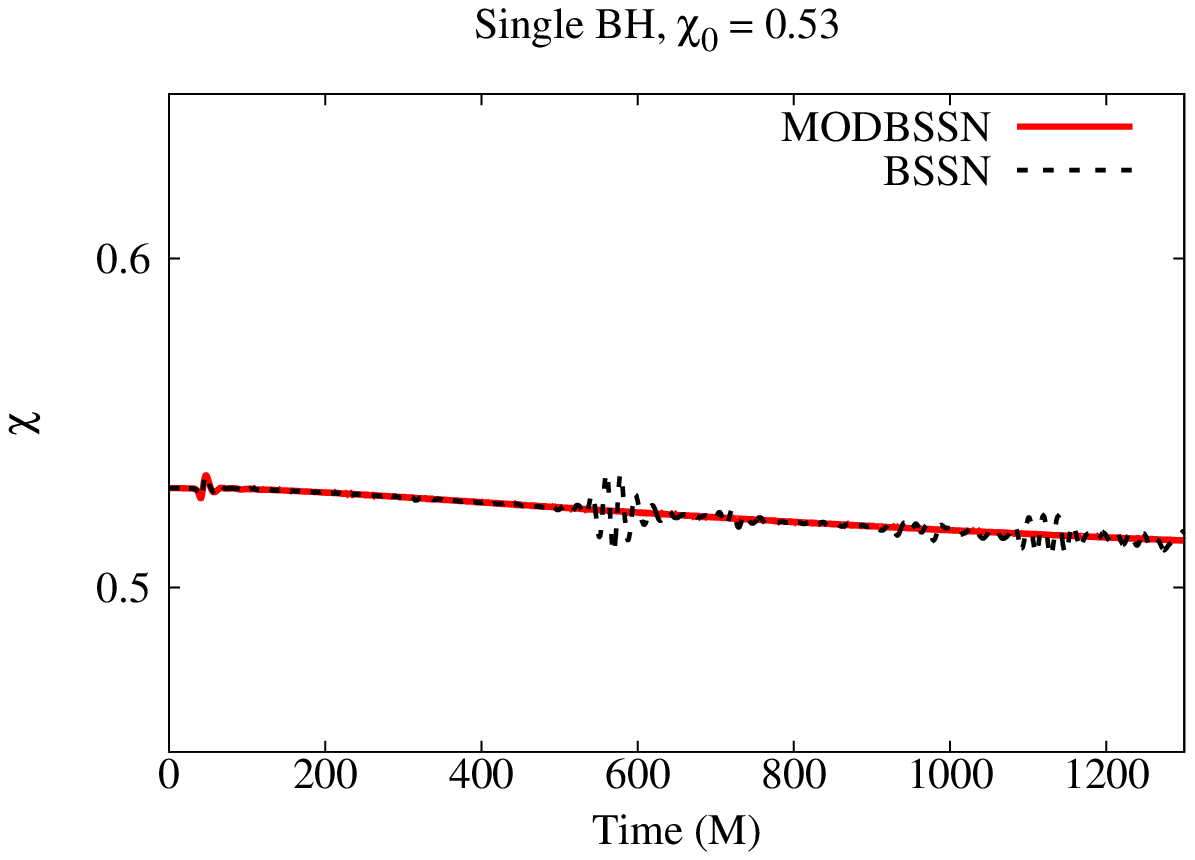}&
\includegraphics[width=\columnwidth]{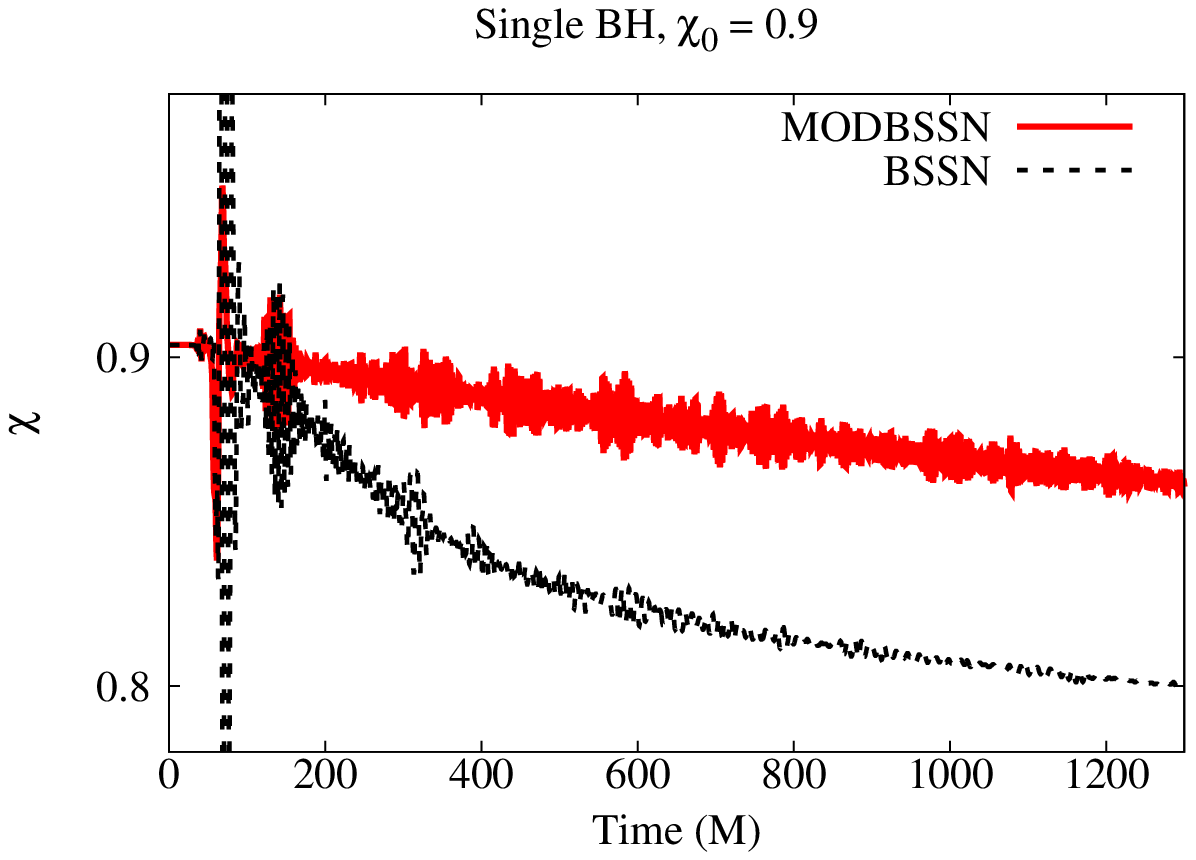}
\end{tabular}
\caption{Dimensionless spin parameter $\chi$ as a function of time for
$\chi_0=0.53$ (left column) and $\chi_0=0.9$ (right column) in the single
black hole evolutions.
The modified BSSN formulation (solid red line, marked as MODBSSN) is shown
to produce less noise in the spurious radiation as well as preserving $\chi$
better than the {\it traditional} BSSN formulation (dashed line, marked as
 BSSN) in the higher spin case.}
\label{fig2}
\end{figure*}
\begin{figure*}[thbp]
\begin{tabular}{rl}
\includegraphics[width=0.98\columnwidth]{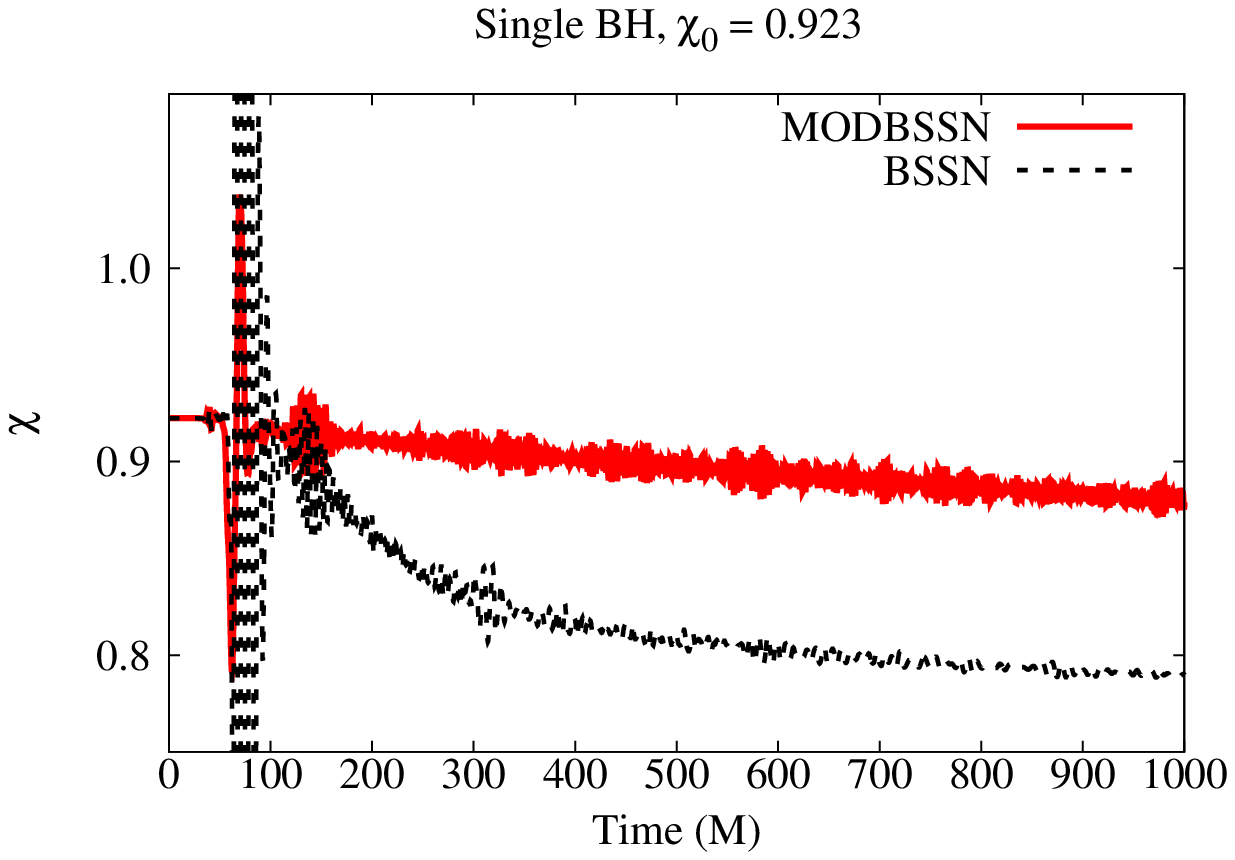}&
\includegraphics[width=\columnwidth]{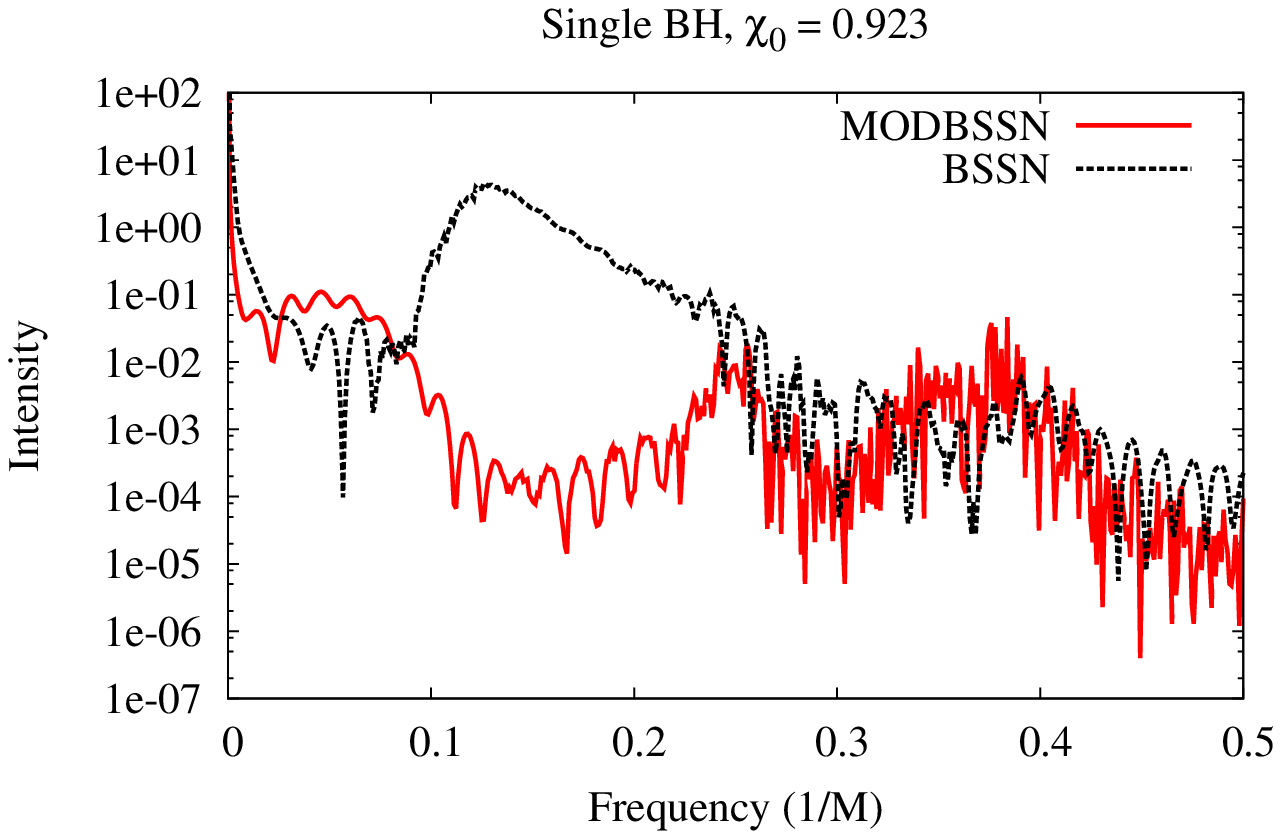}
\end{tabular}
\caption{Left: Dimensionless spin parameter $\chi$ as a function of time
for $\chi_0=0.923$ in the single black hole evolution. The modified
BSSN formulation (solid red line, marked as MODBSSN) is shown to
preserve $\chi$ better than the traditional BSSN formulation (dashed
line, marked as BSSN) in this near extreme case. Right: Power spectrum of
the corresponding data plotted in the left panel.}
\label{nfig1}
\end{figure*}
\subsection{Single black hole tests}\label{sec::IIIA}
In this subsection, we test our modifications in spinning single black
hole (SBH) cases with respective initial dimensionless spin parameters
$\chi_0\equiv{J_0}/{M_0^2}=0.53$, $0.9$, and $0.923$.
To generate these sets of puncture initial data with unit ADM mass,
the bare mass and the spin parameter in the $z$-direction are set to be
$m=0.872335$, $0.3528$, and $0.215898$ and $s_z = 0.53$, $ 0.45$, and
$0.472466$ respectively, as the input for the TwoPuncture solver.
Note that $\chi_0=0.923$ is nearly the maximal spin that the conformally flat
Bowen-York initial data can achieve \cite{dain2008extra}.
The global quantities such as the ADM mass $M$ and the angular momentum $J$ are
calculated with the surface integrals at $R=50M$, as described in \cite{czyy08}.

The dimensionless spin parameter $\chi\equiv{J}/{M^2}$ for the single black
hole simulation is shown in Fig.~\ref{fig2} and Fig.~\ref{nfig1}.
For the cases with the traditional BSSN formulation, it shows that the result
is consistent with the Fig.~4 of \cite{MarTicBru08}.
It is also clear in the figures that the proposed modifications greatly reduce
the overall noise level, and diminish the fluctuation before $t=200$ for the
high-spin case.
The curves of $\chi$ with the modified BSSN formulation show less decay in
each case.
For the lower spin case, $\chi_0=0.53$, the curve for the modified BSSN
formulation (red solid lines) is basically the same as the one for the
traditional BSSN formulation (dashed lines).
As the spin becomes higher, the loss of the angular momentum is more severe.
For the $\chi_0=0.9$ case, the dimensionless spin drops more than $11\%$ to
$0.7986$ at $t=1200$ in the traditional BSSN formulation,
compared to the modified one in which $\chi$ drops less than $5\%$.
This result indicates that the modified BSSN formulation is capable to
conserve the angular momentum better than the traditional BSSN,
especially for the high spin cases, i.e., $\chi_0>0.75$.
\begin{figure*}[thbp]
\begin{tabular}{rl}
\includegraphics[width=\columnwidth]{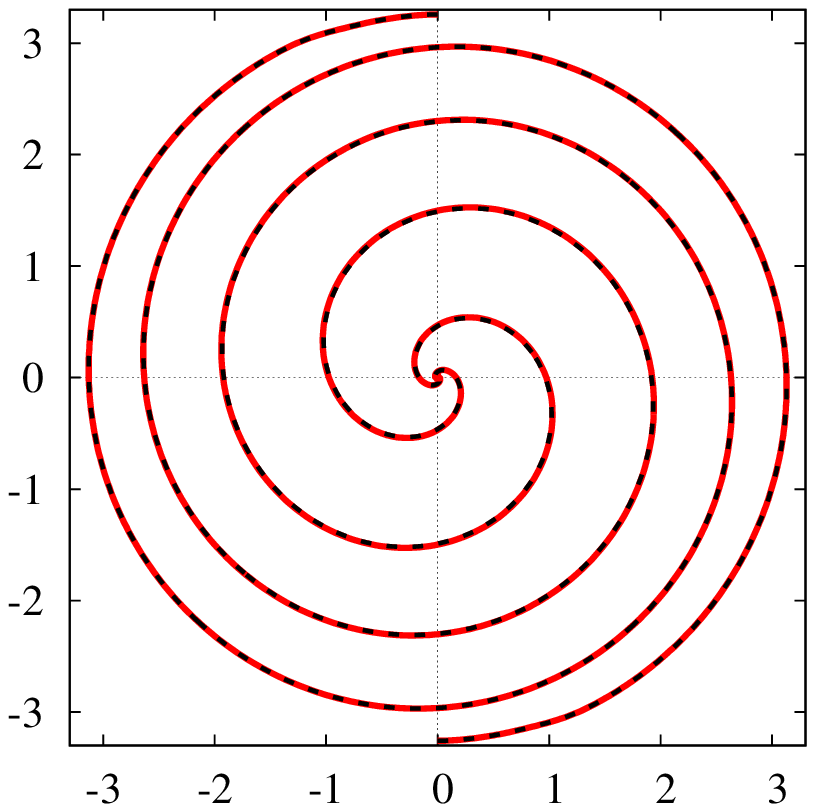}&
\includegraphics[width=\columnwidth]{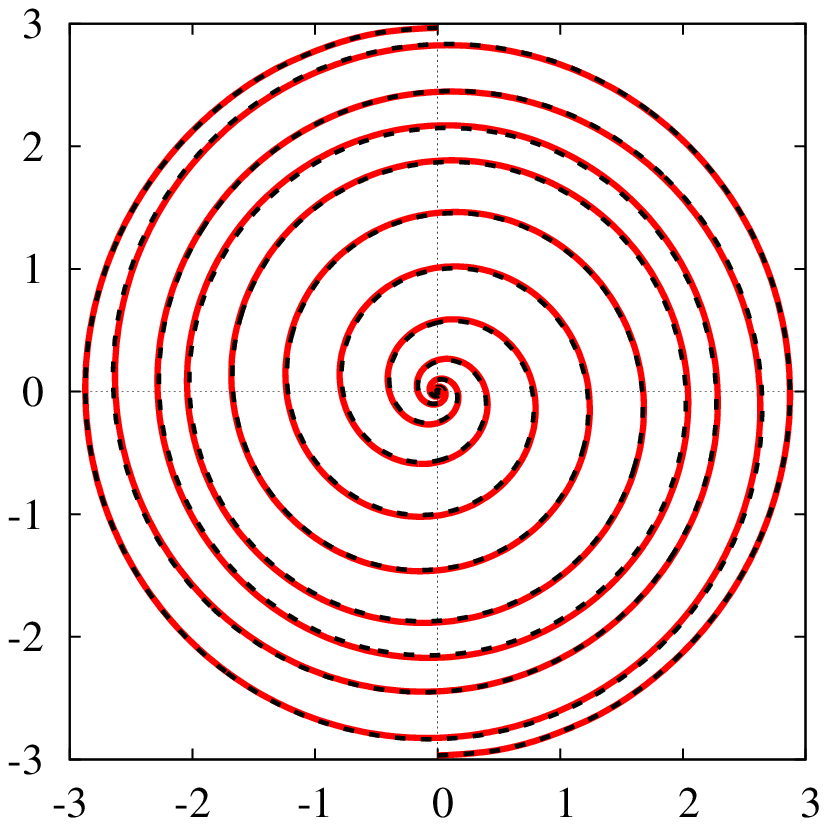}
\end{tabular}
\caption{The puncture trajectory of the binary black hole evolution with
initial $\chi_i=0$ (left) and $\chi_i=0.9$ (right).
The comparisons between the result with the traditional BSSN formulation
(dashed line) and that with the modified BSSN formulation (solid red line)
are shown.}
\label{fig4}
\end{figure*}
\begin{figure*}[thbp]
\begin{tabular}{rl}
\includegraphics[width=\columnwidth]{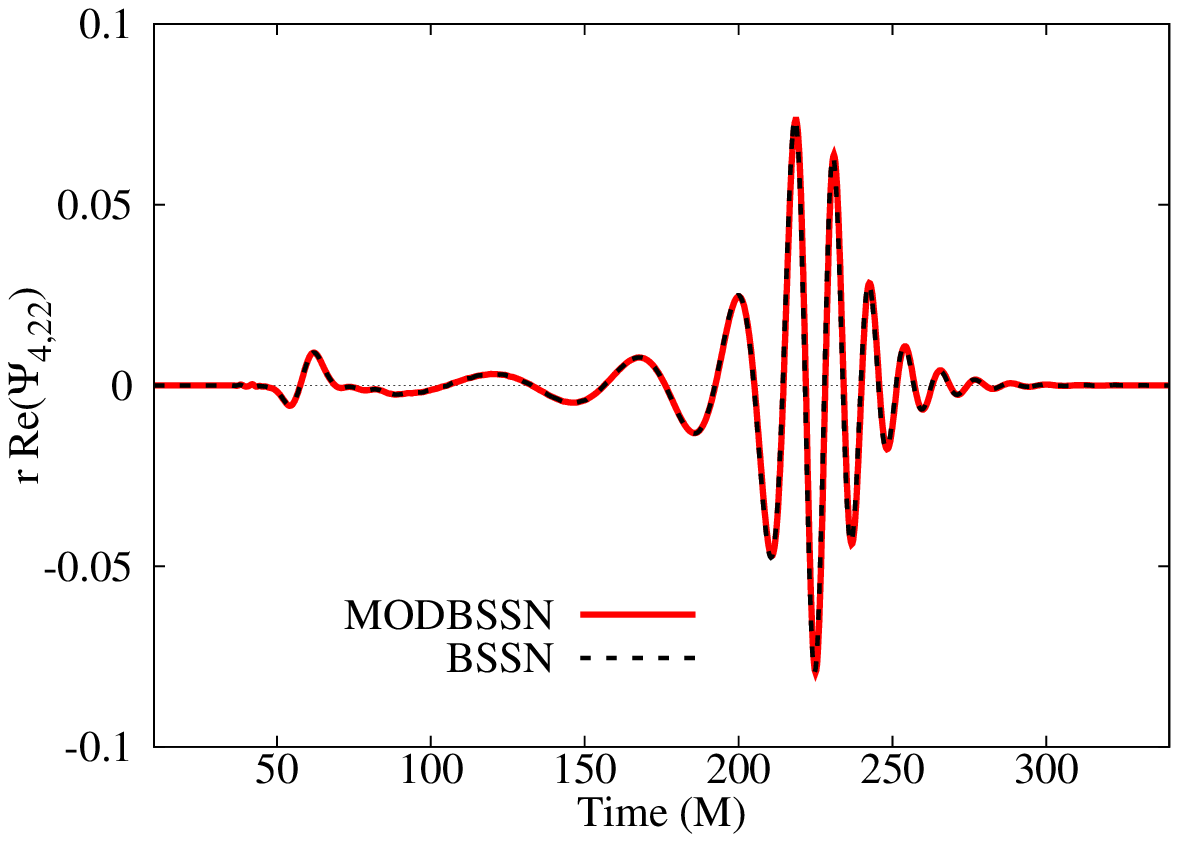}&
\includegraphics[width=\columnwidth]{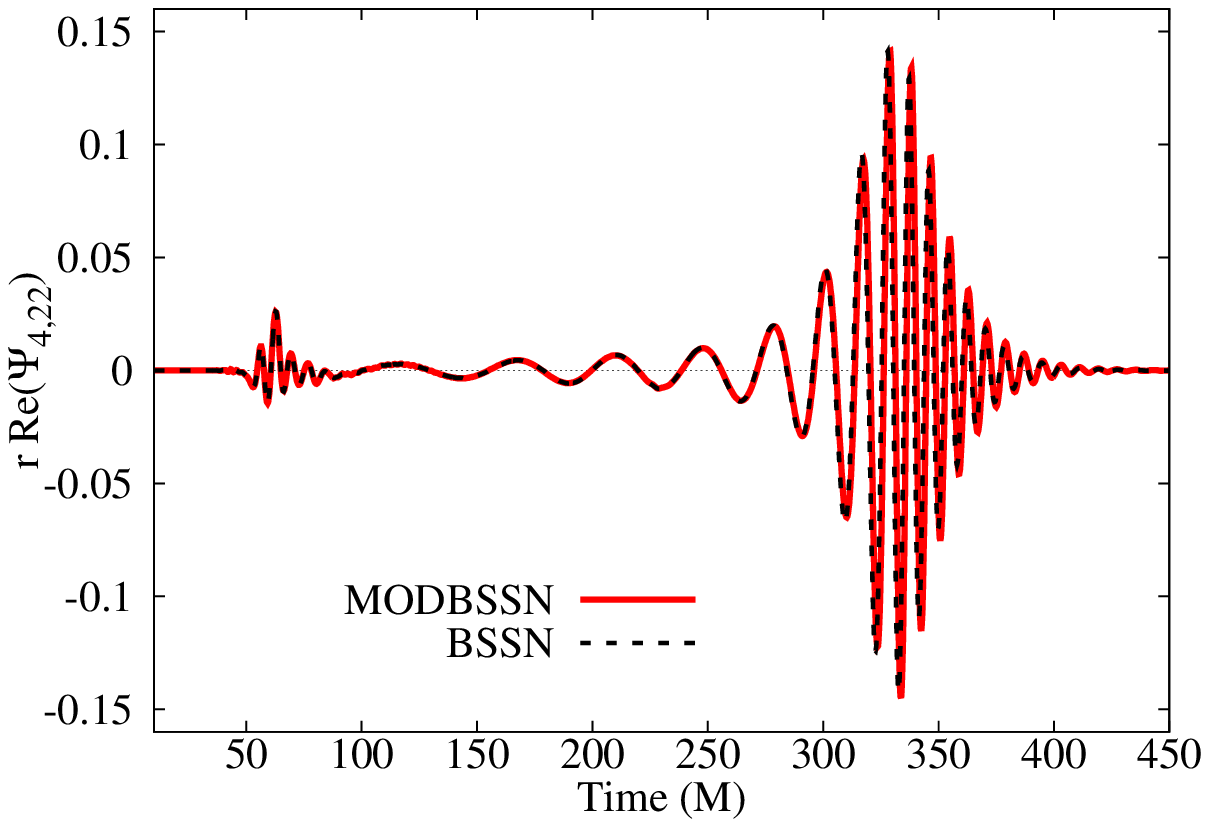}
\end{tabular}
\caption{The $\ell=2$, $m=2$ mode of the Newman-Penrose scalar $\Psi_4$
calculated at $R=50M$, with initial $\chi_i=0$(left) and $\chi_i=0.9$(right).
This mode is the major component of gravitational radiation during the merger
in the binary black hole case with spin parallel to the orbital angular
momentum.
Only the real part of $\Psi_4$ is shown here.
These two waveforms are almost identical for the traditional BSSN formulation
(dashed line, marked as BSSN) and the modified one(solid red line, marked as
MODBSSN).
The binary composed of higher-spin hole spend more inspiral period due to the
spin hang-up effect.}
\label{fig5}
\end{figure*}
\begin{figure*}[thbp]
\begin{tabular}{rl}
\includegraphics[width=\columnwidth]{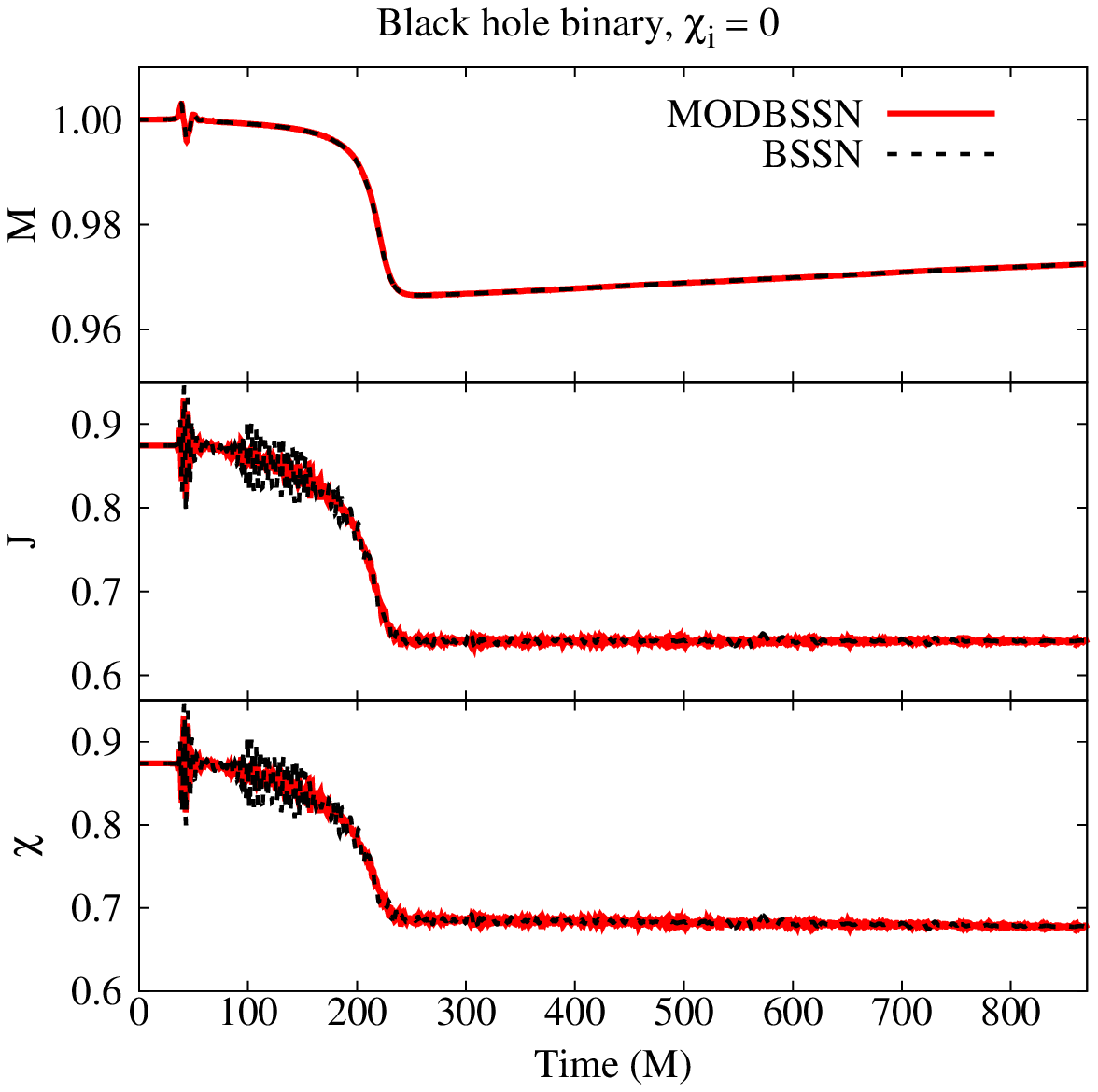}&
\includegraphics[width=\columnwidth]{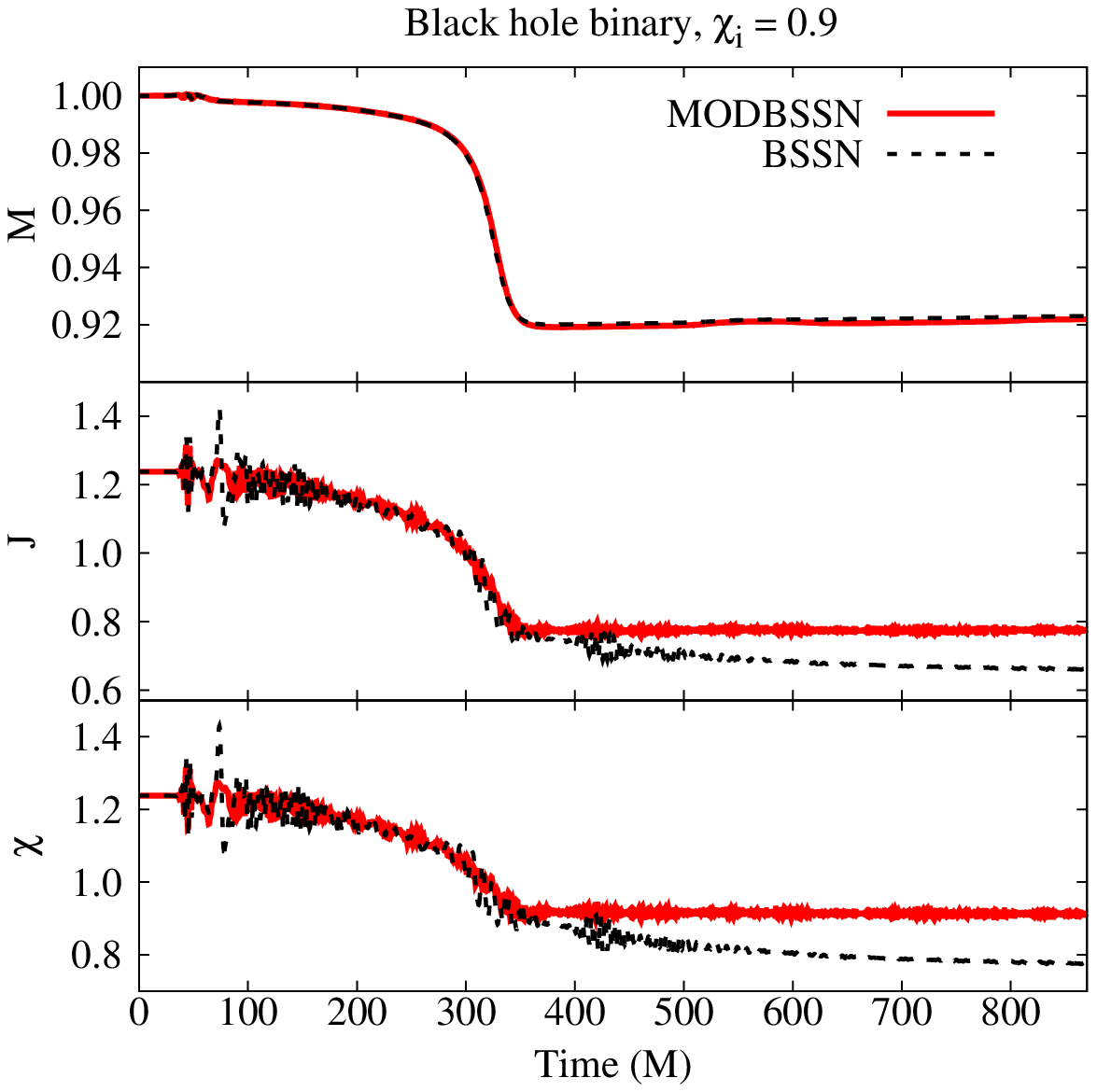}
\end{tabular}
\caption{ADM mass $M$, angular momentum $J$, and dimensionless
spin parameter $\chi$ as functions of time for the initial $\chi_i=0$
(left column) and the $\chi_i=0.9$ (right column) in the binary black
hole evolutions.
The modified BSSN formulation (solid red line, marked as MODBSSN) has better
control than the traditional BSSN(dashed line) on the early noise level and
the spin drop after merger for the higher spin case.}
\label{fig6}
\end{figure*}

The time evolution of the dimensionless spin is shown in the left panel of
Fig.~\ref{nfig1} for
the near-extreme single black hole with $\chi_0=0.923$, which is nearly the
maximal value that the conformally flat Bowen-York data can achieve.
The dimensionless spin drops about $15\%$ to $0.79$ at $t=1000$ in the
traditional BSSN formulation, compared to the modified case in which the change
of $\chi$ is less than $5\%$.
It shows that the modified BSSN formulation is more effective in conserving
the angular momentum over the traditional BSSN formulation, even in the
fast-spinning SBH case.

It is interesting to study the different effect on the simulations between
the $5^{\rm th}$-order KO dissipation and Modification M3.
From a naive observation on Fig.~\ref{fig2} and the left panel of
Fig.~\ref{nfig1}, we found that the KO method is good at eliminating
relatively higher frequency numerical noise.
It can be seen that the result in right panel of Fig.~\ref{fig2} for the
$\chi_0=0.9$ case with the traditional BSSN formulation (dashed line) is
smoother than its counterpart with the modified BSSN formulation
(red solid line), despite the spin drop in the former one.
In contrary, the lower frequency numerical noise appearing in the traditional
BSSN formulation is diminished significantly with the modified BSSN
formulation.
This can also be seen in the left panel of Fig.~\ref{fig2} wherein the two
major fluctuations at $t\approx 580$ and $t\approx 1150$ with the traditional
BSSN formulation (dashed lines) disappear with the modified BSSN formulation.
It also can be seen in the right panel wherein the severe fluctuations before
$t\approx 150$ with the traditional BSSN formulation is effectively
suppressed with the modified BSSN formulation.
It has similar behavior in the left panel of Fig.~\ref{nfig1}.

To understand this phenomenon better, a Fourier analysis method is applied to
the $\chi_0=0.923$ single black hole case.
In the right panel of Fig.~\ref{nfig1}, we show the corresponding power
spectrum of the data in the left panel.
From this power spectrum, we can see that the KO method in the traditioinal
BSSN formulation only dissipates some high-frequency ($f\sim 0.34-0.43$) noise
better.
For the noise in the most other frequency, the M3 method in the modified BSSN
formulation is much more efficient in dissipation.
This difference results in the different behavior in the left panel as we can
see.
The line for traditional BSSN formulation has larger amplitude oscillations
with the intermediate frequency.
The similar results can be seen in Fig.~\ref{fig2}.
The above result indicates that the KO dissipation and M3 suppress
the numerical noise in different frequency ranges.
It is noted that usage of M3 does not introduce any artificial
dissipation and thus the field equation of ${\tilde A}_{ij}$ is analytically
equivalent to the original.

\subsection{Binary black hole tests}\label{sec::IIIB}
In this subsection, we apply our modifications to the equal-mass black hole
binary.
Each black hole in the binary has the spin aligned with the orbital angular
momentum and the dimensionless spin parameter $\chi_i=0.9$ initially.
As the reference, we also run an equal-mass BBH with $\chi_i=0$ for each black
hole.
The initial parameters for each hole as the input of the TwoPuncture solver
are listed in Table \ref{bbh_par}.
\begin{table}[thbp]
\caption{Parameters for the binary black hole puncture initial data}
\centering
\begin{tabular}{ccc}
\hline\hline
$\chi_i$&$0$&$0.9$\\
\hline
bare mass&$0.483$&$0.1764$\\
$\vec r$&$\pm3.257\hat{\bf y}$&$\pm 2.966\hat{\bf y}$\\
$\vec p$&$\mp 0.133\hat{\bf x}$&$\mp 0.12616\hat{\bf x}$\\
$\vec s$&$0$&$0.225\hat{\bf z}$\\
\hline
\end{tabular}
\label{bbh_par}
\end{table}

Firstly we would like to check if our modifications give any changes in these
well-tested BBH cases, compared to the traditional BSSN formulation.
Figures \ref{fig4} and \ref{fig5} show the almost identical puncture
trajectories and the $\ell=2$, $m=2$ mode of the Newman-Penrose
scalar $\Psi_4$ at $R=50M$ for the cases of $\chi_i=0$ and $\chi_i=0.9$.
The $\ell=2$, $m=2$ mode gives the major component of gravitational radiation
during the merger in the case of the binary black hole with spin parallel to the
orbital angular momentum.
Here we only show the real part of $\Psi_4$.
This result is expected in developing new modifications since all of these
formulations are analytically equivalent to Einstein's field equations and
should give same physics.

The ADM mass $M$, the angular momentum $J$, and the dimensionless spin
parameter $\chi$ are shown in Fig.~\ref{fig6} for the BBH cases with the
initial dimensionless spin $\chi_i=0$ (left) and $\chi_i=0.9$ (right).
After the merger at $t=250$ in the $\chi_i=0$ BBH case, $M$ and $J$ decrease
by $4\%$ and $27\%$ respectively due to the gravitational radiation.
And the dimensionless spin parameter after the merger is $\chi=0.68$ at $t=250$
to $\chi=0.67$ at $t>800$.
Thus $\chi$ decreases less than $2\%$ after $t=250$ until the end of simulation.
It also shows that the modified and traditional BSSN formulations give the
same result (in the left panel) in the initially slowly spinning BBH case.
In the $\chi_i=0.9$ case, after the merger at $t=350$, the gravitational
radiation decreases the value of $M$ and $J$ by $8\%$ and $39\%$ respectively.
As shown in the right panel for the $\chi_i=0.9$ case, $\chi$ in the
traditional BSSN formulation decreases more than $10\%$ from $t=350$ to $800$
($\chi\approx 0.75$ in our extended run for $t>1900$).
This result from the traditional BSSN formulation is consistent with the
discovery in \cite{MarTicBru08} in which the final $J$ will decay considerably
for $\chi_i\ge 0.75$.
For the $\chi_i=0.9$ case with the modified BSSN formulation,
$\chi$ decreases only by $1\%$ from $t=350$ to $800$.
The decrease is still less than $2\%$ in the extended run for $t>1900$.
The results in the BBH cases, combined with that in the SBH cases, indicate
clearly that our modifications handle the highly spinning black holes much
better than the traditional BSSN formulation, while yielding the same
results as in traditional BSSN formulation in the slow spinning black
hole cases.

\subsection{Numerical Convergence}
\begin{figure}[thbp]
\includegraphics[width=\columnwidth]{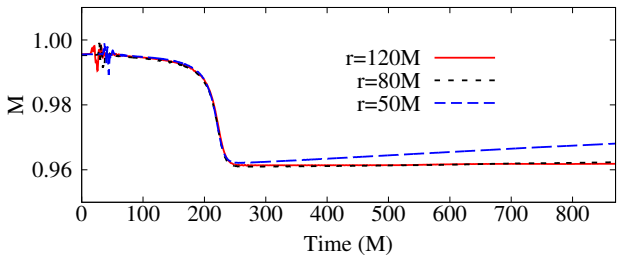}
\caption{
Effect of the extraction radius on the ADM mass integration.
The plot corresponds to the spinless binary black hole case.
The $r$'s in the legend are the extraction radii used in this case.}
\label{nfig3}
\end{figure}
\begin{figure*}[thbp]
\begin{tabular}{rl}
\includegraphics[width=\columnwidth]{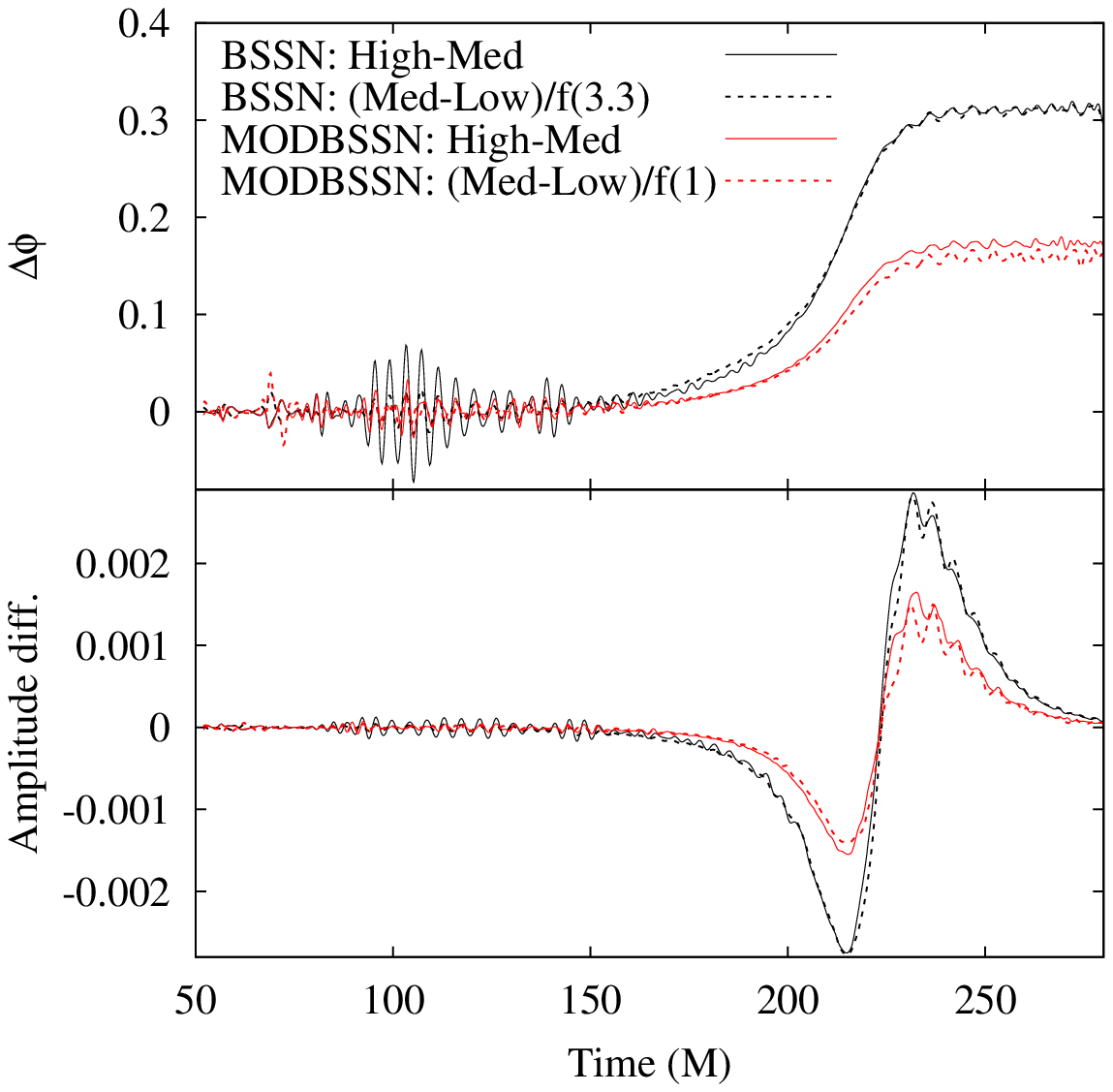}&
\includegraphics[width=\columnwidth]{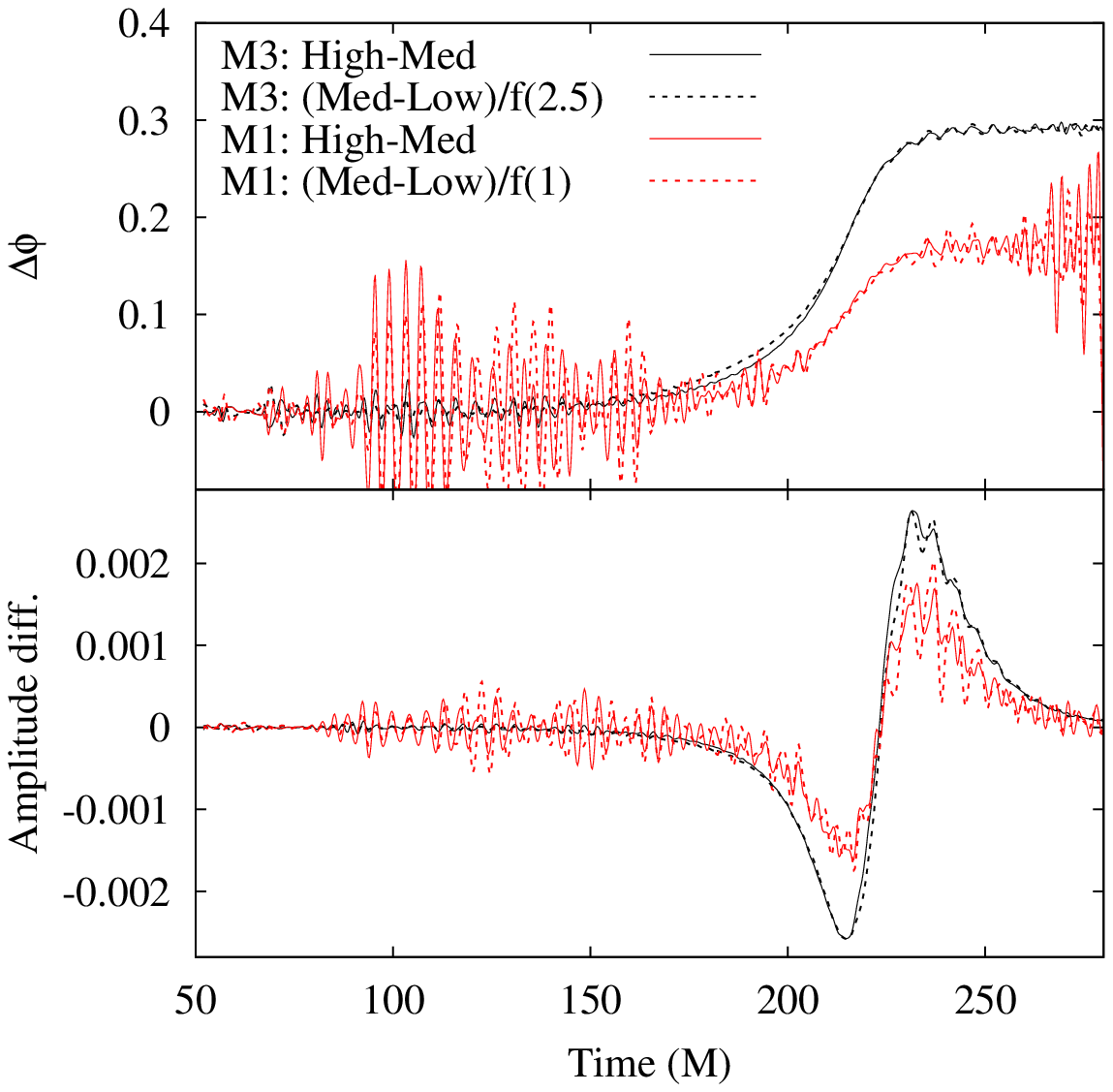}
\end{tabular}
\caption{
Convergence of gravitational wave for the $\chi_i=0$ binary black hole
case.
The left-side panels show the phase differences and the amplitude differences
of $\Psi_4$ respectively between the high and medium resolutions (solid line),
and the medium and low resolutions (dotted line), with both the traditional
BSSN formulation (BSSN, black line) and the modified one (MODBSSN, red line).
Here we use $f(p)$ to denote the factor of order $p$.
The right-side panels show the phase differences and the amplitude differences
of $\Psi_4$ respectively between the high and medium resolutions (solid line),
and the medium and low resolutions (dotted line), with the traditional
BSSN formulation$+$Modification M3 only (M3, black line) and the traditional
one$+$Modification M1 only (M1, red line).}
\label{nfig2}
\end{figure*}
For both the gravitational wave extraction and the calculation of the global
quantities, they are numerically integrated on the  sphere of radius $r=50M$.
This finite radius for integration could affect the accuracy of the amplitude
of the gravitational waveform $\Psi_4$.
However, the effect from the integration sphere of finite radius should be
roughly the same with either the traditional BSSN formulation or the modified
one on any case.
Since we are only concerned about the relative difference between these two
formulations, the effect from the extraction radius becomes unimportant.
For the global quantities, e.g., the ADM mass and the angular momentum,
the integration sphere of finite radius, e.g., $r=50M$, may result in some
artificial drift as shown in Fig.~\ref{fig6}.
Nevertheless, when an integration sphere with larger radius is applied to
the case, such kind of drift diminishes.
Here we use the spinless binary black hole case as an example in
Fig.~\ref{nfig3} for illustration.
According to Fig.~\ref{nfig3}, the result with $r=50M$ are basically same as
the ones with $r=80M$ and $r=120M$ during the merger phase.
The drift in the case with $r=50M$ only shows during the ringdown stage.
And the drift can be easily diminished with larger radii, e.g., $r=120M$ in
Fig.~\ref{nfig3}.
However, in order to compare our result with the one in \cite{MarTicBru08}
closely, in this work we still take the radius $r=50M$, same as used in
\cite{MarTicBru08}.

The physical boundary used in simulations may also affect the gravitational
wave and the calculation of the global quantities.
In order to investigate such possible effects, we have tested the simulations
with farther boundaries.
And our results show that the effect from the boundary condition is ignorable
in the current work.

According to the arguments in Sec.~\ref{sec::II}, analytically we do not expect
Modification M3 to affect the convergence of the system,
and we {\it do} expect that Modification M1 definitely affects the system's
numerical convergent behavior due to the switch character of the step function
in Eq.~(\ref{dtG}).
Here we would like to both check the convergence order of the system with
the modified BSSN formulation and verify these arguments numerically.

Firstly, we show the system's convergence with our modifications in the
left-side panels of Fig.~\ref{nfig2}, compared with the one with the
traditional BSSN formulation.
Here we use the $\chi_i=0$ binary black hole  case as an example for
demonstration.
In the two plots we study the convergence of the phase and the amplitude
for the gravitational wave respectively.
According to the plots, the traditional BSSN formulation results in overall
$3.3^{\rm th}$-order convergence for the system in both the phase and amplitude
of $\Psi_4$, which is roughly consistent with the ideal convergence of
fourth-order with the numerical method used in this work.
On the other hand, it shows in the panels that the modified one results in
only first-order convergence for the system.
Since we already expect that some of our modifications will affect the
convergence of the system, the result is understandable, although the order of
convergence is still considered low.
And we can see from Fig.~\ref{nfig2} that the numerical error with the
modified BSSN formulation is much smaller than the one with the
traditional BSSN formulation, especially in the lower resolutions.
This merit for the modified BSSN formulation could compensate for its
disadvantage of having lower order convergence.
And the convergence behavior showed in Fig.~\ref{nfig2} is general for all
the cases we have done in this work.

Secondly, we would like to confirm the theoretical analysis that it is
Modification M1, not M3, in the modified BSSN formulation which majorly affects
the convergence order.
By using again the $\chi_i=0$ binary black hole case as an example, we show the
result in the right-side panels of Fig.~\ref{nfig2}.
When we apply the traditional BSSN formulation $+$ M1 to the case,
the resulted convergence order is first-order, which is roughly the same as
the convergence order for the case with the whole modified BSSN formulation.
Meanwhile, when we apply the traditional BSSN formulation $+$ M3 to the case,
the resulted convergence order is $2.5^{\rm th}$-order, which is a little
lower than the case with the pure traditional BSSN formulation,
but quite higher than the one with the modified one..
The result tells that M1 is the key modification which affects majorly the
convergence behavior of the system, as we expected.
However, it also shows that M3 lowers minorly the convergence order of
the system.
This indicates that the convergence order of M3 might not be {\it numerically}
as good as the expectation from our analytical argument
\footnote{
It is quite possible that the numerical convergence order with
Modification M3 is lower than the one with pure traditional BSSN formulation
due to some numerical residual from the finite-differencing in M3, although
its convergence order is designed to be higher than analytically.
However, in the modified BSSN formulation, its convergence order turns out to
be irrelevant as long as its order is higher than the convergence order of M1.
}.
In conclusion, the cases with our modifications have first-order convergence,
which is lower than the one with the traditional BSSN formulation,
but our modifications give more accurate result than the traditional
BSSN formulation at a given resolution.
And Modification M1 is the key factor in our modifications to affect the
convergence behavior.
\section{Discussion and summary}\label{sec::IV}
In this work, we applied our modifications of the BSSN formulation to study
the total angular momentum conservation issue in black hole evolutions
with the standard Bowen-York puncture initial data.
We found that the non-negligible loss of angular momentum for highly spinning
black holes mentioned in \cite{MarTicBru08} can be greatly cured with
our modifications.
The improvements are obvious for near-extreme cases, as in the
SBH case shown in Fig.~\ref{fig2} and Fig.~\ref{nfig1} and the BBH case
in Fig.~\ref{fig6}.
It has also been shown in the previous section that the
modified BSSN formulation does not introduce any unphysical effects.
Improving the conservation of the angular momentum usually leads to
certain improvement on the accuracy of the results in black hole evolutions.
Therefore we expect that our modifications will provide better performance in
black hole evolution simulations than the traditional BSSN scheme.

Modification M1 is the most important to the
conservation of the angular momentum since the field equation of the
conformal connection function ${\tilde\Gamma}{}^i$ is closely related to
the (angular) momentum vector.
We find that Eq.~(\ref{dtG}) and setting $\xi=1$ gives quite robust and stable
runs. But due to the switch character of M1 and the possible sign fluctuation of the numerical
value of its argument when the argument is close to zero,
the step function is sensitive to the resolution used in simulations.
So this modification affects majorly the numerical
convergence order of the modified BSSN formulation.
M2 is able to enhance the hyperbolicity of the system, especially for the
evolution of ${\tilde\Gamma}{}^i$.
However, its mechanism and the optimal choice of $\sigma$ need further
investigations.

Instead of the KO dissipation method used in the traditional BSSN formulation,
Modification M3 is used in the modified BSSN formulation in this work.
We can see in Sec.~\ref{sec::III} that M3 is able to diminish effectively
some intermediate frequency noise with larger amplitude;
while the KO dissipation is good at eliminating the higher frequency
numerical noise.
To some extent, Modification M3 is complementary to the KO method in
dissipating numerical error.
However, the advantage of M3 is that it comes from the derivative of the
momentum constraint.
Thus, applying M3 to the BSSN formulation is always legitimate and safe
as long as the momentum constraint holds.
In contrast, the application of the KO method is not always safe since
it is an artificial addition to the field equation, although it is convenient
and effective in numerical relativity.
It is possible that usage of the KO method leads to deviations of the
numerical result from the solution hypersurface, especially when the result
is sensitive to the initial data.
It will be a good idea to use both Modification M3 and the KO method in
dissipating the numerical error in the future simulations.

As mentioned at the beginning of Sec.~\ref{sec::II},
in order to enforce the algebraic constraints of the unimodular determinant
 of $\tilde\gamma_{ij}$, and of $\tilde A_{ij}$ being traceless,
the numerical values of
$\tilde\gamma_{ij}$ and $\tilde A_{ij}$ are replaced with
$\tilde\gamma_{ij}\rightarrow\tilde\gamma^{-1/3}\tilde\gamma_{ij}$,
$\tilde A_{ij}\rightarrow\tilde A_{\langle ij\rangle}$ after every time step
in the traditional BSSN formulation.
However, the modification
\begin{align}
{\tilde\gamma}_{zz}&\rightarrow\frac{1+{\tilde\gamma}_{yy}{\tilde\gamma}_{xz}^2
-2{\tilde\gamma}_{xy}{\tilde\gamma}_{yz}{\tilde\gamma}_{xz}
+{\tilde\gamma}_{xx}{\tilde\gamma}_{yz}^2}{{\tilde\gamma}_{xx}
{\tilde\gamma}_{yy}-{\tilde\gamma}_{xy}^2},\label{gzz}\\
{\tilde A}_{yy}&\rightarrow\frac{{\tilde A}_x{}^x+{\tilde A}_z{}^z
+{\tilde A}_{xy}{\tilde\gamma}^{xy}+{\tilde A}_{yz}{\tilde\gamma}^{yz}}
{{\tilde\gamma}^{yy}},\label{Ayy}
\end{align}
 is employed instead in \cite{yhlc12} to enforce the two constraints.
The results in \cite{yhlc12} have shown that this modification gains
better stability compared to the traditional BSSN formulation.
Differing with \cite{yhlc12}, in this work we use the traditional recipe of
$\tilde\gamma_{ij}\rightarrow\tilde\gamma^{-1/3}\tilde\gamma_{ij}$,
$\tilde A_{ij}\rightarrow\tilde A_{\langle ij\rangle}$ instead of applying
the modification Eqs.~(\ref{gzz}) \& (\ref{Ayy}).
This is because the denominators in Eqs.~(\ref{gzz},\ref{Ayy}) can be very
small near the singularities and thus the numerical values of the replaced
${\tilde\gamma}_{zz}$ and ${\tilde A}_{yy}$ can have unexpected fluctuation
which can crash the code.
However, this modification can still be applied to the BSSN formulation
if there is no singularity or if an excision method is
used in the black hole evolution simulations.

In the modifications, we introduce some terms related to the spacial
resolution used in the numerical simulation.
These terms reduce the convergence order from fourth-order to first-order
in our implementation. But compared with the traditional BSSN formulation,
the numerical error resulted in the modified BSSN formulation is still
obviously smaller than the one from the traditional BSSN formulation
with a reasonably fine resolution.

In this work, we demonstrated the simulations with the modified BSSN
formulation in which the angular momentum conservation is better than in the
traditional BSSN formulation.
Thus this modified BSSN formulation should improve the accuracy in the
punctured black hole evolutions.
Our modifications are imposed on the field equations of the physical variables
${\tilde\gamma}_{ij}$, ${\tilde A}_{ij}$ and ${\tilde\Gamma}{}^i$, instead of
the gauge variables $\alpha$ and $\beta^i$.
Therefore, we expect that the modified BSSN formulation can be applied
generally to various scenarios to give improved results in numerical
relativity.
\section*{Acknowledgments}
This work was supported in part by the National Science Council under Grants
No.~NSC102-2112-M-006-014-MY2,
and by the Headquarters of University Advancement at the National Cheng Kung
University, which is sponsored by the Ministry of Education, Taiwan, ROC.
Z.~Cao was supported by the NSFC (No.~11375260).
We are grateful to the National Center for High-performance Computing
for the use of their computer time and facilities. We are also
grateful to the Academia Sinica Computing Center for providing
computing resource; and to Peter Diener for helpful discussions.
HJY thanks the hospitality of ASIAA for hosting his visit from July 2014
to February 2015.
\bibliography{ref}

\begin{thebibliography}{39}
\expandafter\ifx\csname natexlab\endcsname\relax\def\natexlab#1{#1}\fi
\expandafter\ifx\csname bibnamefont\endcsname\relax
  \def\bibnamefont#1{#1}\fi
\expandafter\ifx\csname bibfnamefont\endcsname\relax
  \def\bibfnamefont#1{#1}\fi
\expandafter\ifx\csname citenamefont\endcsname\relax
  \def\citenamefont#1{#1}\fi
\expandafter\ifx\csname url\endcsname\relax
  \def\url#1{\texttt{#1}}\fi
\expandafter\ifx\csname urlprefix\endcsname\relax\def\urlprefix{URL }\fi
\providecommand{\bibinfo}[2]{#2}
\providecommand{\eprint}[2][]{\url{#2}}

\bibitem[{\citenamefont{Yo et~al.}(2012)\citenamefont{Yo, Lin, and
  Cao}}]{yhlc12}
\bibinfo{author}{\bibfnamefont{H.-J.} \bibnamefont{Yo}},
  \bibinfo{author}{\bibfnamefont{C.-Y.} \bibnamefont{Lin}}, \bibnamefont{and}
  \bibinfo{author}{\bibfnamefont{Z.}~\bibnamefont{Cao}},
  \bibinfo{journal}{Phys. Rev. D} \textbf{\bibinfo{volume}{86}},
  \bibinfo{pages}{064027} (\bibinfo{year}{2012}),
  \urlprefix\url{http://link.aps.org/doi/10.1103/PhysRevD.86.064027}.

\bibitem[{\citenamefont{Pretorius}(2005)}]{pref05}
\bibinfo{author}{\bibfnamefont{F.}~\bibnamefont{Pretorius}},
  \bibinfo{journal}{Physical review letters} \textbf{\bibinfo{volume}{95}},
  \bibinfo{pages}{121101} (\bibinfo{year}{2005}).

\bibitem[{\citenamefont{Campanelli et~al.}(2006)\citenamefont{Campanelli,
  Lousto, Marronetti, and Zlochower}}]{NR06a}
\bibinfo{author}{\bibfnamefont{M.}~\bibnamefont{Campanelli}},
  \bibinfo{author}{\bibfnamefont{C.~O.} \bibnamefont{Lousto}},
  \bibinfo{author}{\bibfnamefont{P.}~\bibnamefont{Marronetti}},
  \bibnamefont{and}
  \bibinfo{author}{\bibfnamefont{Y.}~\bibnamefont{Zlochower}},
  \bibinfo{journal}{Physical review letters} \textbf{\bibinfo{volume}{96}},
  \bibinfo{pages}{111101} (\bibinfo{year}{2006}).

\bibitem[{\citenamefont{Baker et~al.}(2006)\citenamefont{Baker, Centrella,
  Choi, Koppitz, and van Meter}}]{NR06b}
\bibinfo{author}{\bibfnamefont{J.~G.} \bibnamefont{Baker}},
  \bibinfo{author}{\bibfnamefont{J.}~\bibnamefont{Centrella}},
  \bibinfo{author}{\bibfnamefont{D.-I.} \bibnamefont{Choi}},
  \bibinfo{author}{\bibfnamefont{M.}~\bibnamefont{Koppitz}}, \bibnamefont{and}
  \bibinfo{author}{\bibfnamefont{J.}~\bibnamefont{van Meter}},
  \bibinfo{journal}{Physical review letters} \textbf{\bibinfo{volume}{96}},
  \bibinfo{pages}{111102} (\bibinfo{year}{2006}).

\bibitem[{\citenamefont{Pan et~al.}(2010)\citenamefont{Pan, Buonanno, Buchman,
  Chu, Kidder, Pfeiffer, and Scheel}}]{PanBuoBuc10}
\bibinfo{author}{\bibfnamefont{Y.}~\bibnamefont{Pan}},
  \bibinfo{author}{\bibfnamefont{A.}~\bibnamefont{Buonanno}},
  \bibinfo{author}{\bibfnamefont{L.~T.} \bibnamefont{Buchman}},
  \bibinfo{author}{\bibfnamefont{T.}~\bibnamefont{Chu}},
  \bibinfo{author}{\bibfnamefont{L.~E.} \bibnamefont{Kidder}},
  \bibinfo{author}{\bibfnamefont{H.~P.} \bibnamefont{Pfeiffer}},
  \bibnamefont{and} \bibinfo{author}{\bibfnamefont{M.~A.}
  \bibnamefont{Scheel}}, \bibinfo{journal}{Phys. Rev. D}
  \textbf{\bibinfo{volume}{81}}, \bibinfo{pages}{084041}
  (\bibinfo{year}{2010}),
  \urlprefix\url{http://link.aps.org/doi/10.1103/PhysRevD.81.084041}.

\bibitem[{\citenamefont{Campanelli et~al.}(2007)\citenamefont{Campanelli,
  Lousto, Zlochower, and Merritt}}]{BHkicksa}
\bibinfo{author}{\bibfnamefont{M.}~\bibnamefont{Campanelli}},
  \bibinfo{author}{\bibfnamefont{C.~O.} \bibnamefont{Lousto}},
  \bibinfo{author}{\bibfnamefont{Y.}~\bibnamefont{Zlochower}},
  \bibnamefont{and} \bibinfo{author}{\bibfnamefont{D.}~\bibnamefont{Merritt}},
  \bibinfo{journal}{Physical Review Letters} \textbf{\bibinfo{volume}{98}},
  \bibinfo{pages}{231102} (\bibinfo{year}{2007}).

\bibitem[{\citenamefont{Lousto and Zlochower}(2008)}]{BHkicksb}
\bibinfo{author}{\bibfnamefont{C.~O.} \bibnamefont{Lousto}} \bibnamefont{and}
  \bibinfo{author}{\bibfnamefont{Y.}~\bibnamefont{Zlochower}},
  \bibinfo{journal}{Physical Review D} \textbf{\bibinfo{volume}{77}},
  \bibinfo{pages}{044028} (\bibinfo{year}{2008}).

\bibitem[{\citenamefont{Zlochower et~al.}(2011)\citenamefont{Zlochower,
  Campanelli, and Lousto}}]{BHkicksc}
\bibinfo{author}{\bibfnamefont{Y.}~\bibnamefont{Zlochower}},
  \bibinfo{author}{\bibfnamefont{M.}~\bibnamefont{Campanelli}},
  \bibnamefont{and} \bibinfo{author}{\bibfnamefont{C.~O.}
  \bibnamefont{Lousto}}, \bibinfo{journal}{Classical and Quantum Gravity}
  \textbf{\bibinfo{volume}{28}}, \bibinfo{pages}{114015}
  (\bibinfo{year}{2011}).

\bibitem[{\citenamefont{Lousto and Zlochower}(2011)}]{BHkicksd}
\bibinfo{author}{\bibfnamefont{C.~O.} \bibnamefont{Lousto}} \bibnamefont{and}
  \bibinfo{author}{\bibfnamefont{Y.}~\bibnamefont{Zlochower}},
  \bibinfo{journal}{Physical Review D} \textbf{\bibinfo{volume}{83}},
  \bibinfo{pages}{024003} (\bibinfo{year}{2011}).

\bibitem[{\citenamefont{Lousto et~al.}(2012)\citenamefont{Lousto, Zlochower,
  Dotti, and Volonteri}}]{BHkickse}
\bibinfo{author}{\bibfnamefont{C.~O.} \bibnamefont{Lousto}},
  \bibinfo{author}{\bibfnamefont{Y.}~\bibnamefont{Zlochower}},
  \bibinfo{author}{\bibfnamefont{M.}~\bibnamefont{Dotti}}, \bibnamefont{and}
  \bibinfo{author}{\bibfnamefont{M.}~\bibnamefont{Volonteri}},
  \bibinfo{journal}{Physical Review D} \textbf{\bibinfo{volume}{85}},
  \bibinfo{pages}{084015} (\bibinfo{year}{2012}).

\bibitem[{\citenamefont{Read et~al.}(2009)\citenamefont{Read, Markakis,
  Shibata, Ury{\=u}, Creighton, and Friedman}}]{ReaMarShi09}
\bibinfo{author}{\bibfnamefont{J.~S.} \bibnamefont{Read}},
  \bibinfo{author}{\bibfnamefont{C.}~\bibnamefont{Markakis}},
  \bibinfo{author}{\bibfnamefont{M.}~\bibnamefont{Shibata}},
  \bibinfo{author}{\bibfnamefont{K.}~\bibnamefont{Ury{\=u}}},
  \bibinfo{author}{\bibfnamefont{J.~D.} \bibnamefont{Creighton}},
  \bibnamefont{and} \bibinfo{author}{\bibfnamefont{J.~L.}
  \bibnamefont{Friedman}}, \bibinfo{journal}{Physical Review D}
  \textbf{\bibinfo{volume}{79}}, \bibinfo{pages}{124033}
  (\bibinfo{year}{2009}).

\bibitem[{\citenamefont{Bernuzzi
  et~al.}(2012{\natexlab{a}})\citenamefont{Bernuzzi, Thierfelder, and
  Br\"ugmann}}]{PhysRevD.85.104030}
\bibinfo{author}{\bibfnamefont{S.}~\bibnamefont{Bernuzzi}},
  \bibinfo{author}{\bibfnamefont{M.}~\bibnamefont{Thierfelder}},
  \bibnamefont{and}
  \bibinfo{author}{\bibfnamefont{B.}~\bibnamefont{Br\"ugmann}},
  \bibinfo{journal}{Phys. Rev. D} \textbf{\bibinfo{volume}{85}},
  \bibinfo{pages}{104030} (\bibinfo{year}{2012}{\natexlab{a}}),
  \urlprefix\url{http://link.aps.org/doi/10.1103/PhysRevD.85.104030}.

\bibitem[{\citenamefont{Bernuzzi
  et~al.}(2012{\natexlab{b}})\citenamefont{Bernuzzi, Nagar, Thierfelder, and
  Br\"ugmann}}]{PhysRevD.86.044030}
\bibinfo{author}{\bibfnamefont{S.}~\bibnamefont{Bernuzzi}},
  \bibinfo{author}{\bibfnamefont{A.}~\bibnamefont{Nagar}},
  \bibinfo{author}{\bibfnamefont{M.}~\bibnamefont{Thierfelder}},
  \bibnamefont{and}
  \bibinfo{author}{\bibfnamefont{B.}~\bibnamefont{Br\"ugmann}},
  \bibinfo{journal}{Phys. Rev. D} \textbf{\bibinfo{volume}{86}},
  \bibinfo{pages}{044030} (\bibinfo{year}{2012}{\natexlab{b}}),
  \urlprefix\url{http://link.aps.org/doi/10.1103/PhysRevD.86.044030}.

\bibitem[{\citenamefont{Hotokezaka et~al.}(2013)\citenamefont{Hotokezaka,
  Kyutoku, and Shibata}}]{PhysRevD.87.044001}
\bibinfo{author}{\bibfnamefont{K.}~\bibnamefont{Hotokezaka}},
  \bibinfo{author}{\bibfnamefont{K.}~\bibnamefont{Kyutoku}}, \bibnamefont{and}
  \bibinfo{author}{\bibfnamefont{M.}~\bibnamefont{Shibata}},
  \bibinfo{journal}{Phys. Rev. D} \textbf{\bibinfo{volume}{87}},
  \bibinfo{pages}{044001} (\bibinfo{year}{2013}),
  \urlprefix\url{http://link.aps.org/doi/10.1103/PhysRevD.87.044001}.

\bibitem[{\citenamefont{McKinney and Blandford}(2009)}]{JetBHa}
\bibinfo{author}{\bibfnamefont{J.~C.} \bibnamefont{McKinney}} \bibnamefont{and}
  \bibinfo{author}{\bibfnamefont{R.~D.} \bibnamefont{Blandford}},
  \bibinfo{journal}{Monthly Notices of the Royal Astronomical Society: Letters}
  \textbf{\bibinfo{volume}{394}}, \bibinfo{pages}{L126} (\bibinfo{year}{2009}).

\bibitem[{\citenamefont{Palenzuela et~al.}(2010)\citenamefont{Palenzuela,
  Lehner, and Liebling}}]{JetBHb}
\bibinfo{author}{\bibfnamefont{C.}~\bibnamefont{Palenzuela}},
  \bibinfo{author}{\bibfnamefont{L.}~\bibnamefont{Lehner}}, \bibnamefont{and}
  \bibinfo{author}{\bibfnamefont{S.~L.} \bibnamefont{Liebling}},
  \bibinfo{journal}{Science} \textbf{\bibinfo{volume}{329}},
  \bibinfo{pages}{927} (\bibinfo{year}{2010}).

\bibitem[{\citenamefont{Kiuchi et~al.}(2010)\citenamefont{Kiuchi, Sekiguchi,
  Shibata, and Taniguchi}}]{KiuSekShi10}
\bibinfo{author}{\bibfnamefont{K.}~\bibnamefont{Kiuchi}},
  \bibinfo{author}{\bibfnamefont{Y.}~\bibnamefont{Sekiguchi}},
  \bibinfo{author}{\bibfnamefont{M.}~\bibnamefont{Shibata}}, \bibnamefont{and}
  \bibinfo{author}{\bibfnamefont{K.}~\bibnamefont{Taniguchi}},
  \bibinfo{journal}{Physical review letters} \textbf{\bibinfo{volume}{104}},
  \bibinfo{pages}{141101} (\bibinfo{year}{2010}).

\bibitem[{\citenamefont{Sekiguchi and Shibata}(2011)}]{SekShi11}
\bibinfo{author}{\bibfnamefont{Y.}~\bibnamefont{Sekiguchi}} \bibnamefont{and}
  \bibinfo{author}{\bibfnamefont{M.}~\bibnamefont{Shibata}},
  \bibinfo{journal}{The Astrophysical Journal} \textbf{\bibinfo{volume}{737}},
  \bibinfo{pages}{6} (\bibinfo{year}{2011}).

\bibitem[{\citenamefont{Shibata and Nakamura}(1995)}]{BSSN95_99a}
\bibinfo{author}{\bibfnamefont{M.}~\bibnamefont{Shibata}} \bibnamefont{and}
  \bibinfo{author}{\bibfnamefont{T.}~\bibnamefont{Nakamura}},
  \bibinfo{journal}{Physical Review D} \textbf{\bibinfo{volume}{52}},
  \bibinfo{pages}{5428} (\bibinfo{year}{1995}).

\bibitem[{\citenamefont{Baumgarte and Shapiro}(1998)}]{BSSN95_99b}
\bibinfo{author}{\bibfnamefont{T.~W.} \bibnamefont{Baumgarte}}
  \bibnamefont{and} \bibinfo{author}{\bibfnamefont{S.~L.}
  \bibnamefont{Shapiro}}, \bibinfo{journal}{Physical Review D}
  \textbf{\bibinfo{volume}{59}}, \bibinfo{pages}{024007}
  (\bibinfo{year}{1998}).

\bibitem[{\citenamefont{Alcubierre and Br{\"u}gmann}(2001)}]{almb01}
\bibinfo{author}{\bibfnamefont{M.}~\bibnamefont{Alcubierre}} \bibnamefont{and}
  \bibinfo{author}{\bibfnamefont{B.}~\bibnamefont{Br{\"u}gmann}},
  \bibinfo{journal}{Physical Review D} \textbf{\bibinfo{volume}{63}},
  \bibinfo{pages}{104006} (\bibinfo{year}{2001}).

\bibitem[{\citenamefont{Yo et~al.}(2002)\citenamefont{Yo, Baumgarte, and
  Shapiro}}]{yhbs02}
\bibinfo{author}{\bibfnamefont{H.-J.} \bibnamefont{Yo}},
  \bibinfo{author}{\bibfnamefont{T.~W.} \bibnamefont{Baumgarte}},
  \bibnamefont{and} \bibinfo{author}{\bibfnamefont{S.~L.}
  \bibnamefont{Shapiro}}, \bibinfo{journal}{Physical Review D}
  \textbf{\bibinfo{volume}{66}}, \bibinfo{pages}{084026}
  (\bibinfo{year}{2002}).

\bibitem[{\citenamefont{Yoneda and Shinkai}(2002)}]{ygsh02}
\bibinfo{author}{\bibfnamefont{G.}~\bibnamefont{Yoneda}} \bibnamefont{and}
  \bibinfo{author}{\bibfnamefont{H.}~\bibnamefont{Shinkai}},
  \bibinfo{journal}{Physical Review D} \textbf{\bibinfo{volume}{66}},
  \bibinfo{pages}{124003} (\bibinfo{year}{2002}).

\bibitem[{\citenamefont{Kiuchi and Shinkai}(2008)}]{kksh08}
\bibinfo{author}{\bibfnamefont{K.}~\bibnamefont{Kiuchi}} \bibnamefont{and}
  \bibinfo{author}{\bibfnamefont{H.-a.} \bibnamefont{Shinkai}},
  \bibinfo{journal}{Physical Review D} \textbf{\bibinfo{volume}{77}},
  \bibinfo{pages}{044010} (\bibinfo{year}{2008}).

\bibitem[{\citenamefont{Tsuchiya et~al.}(2012)\citenamefont{Tsuchiya, Yoneda,
  and Shinkai}}]{ttys12}
\bibinfo{author}{\bibfnamefont{T.}~\bibnamefont{Tsuchiya}},
  \bibinfo{author}{\bibfnamefont{G.}~\bibnamefont{Yoneda}}, \bibnamefont{and}
  \bibinfo{author}{\bibfnamefont{H.-a.} \bibnamefont{Shinkai}},
  \bibinfo{journal}{Physical Review D} \textbf{\bibinfo{volume}{85}},
  \bibinfo{pages}{044018} (\bibinfo{year}{2012}).

\bibitem[{\citenamefont{Brown et~al.}(2012)\citenamefont{Brown, Diener, Field,
  Hesthaven, Herrmann, Mrou{\'e}, Sarbach, Schnetter, Tiglio, and
  Wagman}}]{bjet12}
\bibinfo{author}{\bibfnamefont{J.~D.} \bibnamefont{Brown}},
  \bibinfo{author}{\bibfnamefont{P.}~\bibnamefont{Diener}},
  \bibinfo{author}{\bibfnamefont{S.~E.} \bibnamefont{Field}},
  \bibinfo{author}{\bibfnamefont{J.~S.} \bibnamefont{Hesthaven}},
  \bibinfo{author}{\bibfnamefont{F.}~\bibnamefont{Herrmann}},
  \bibinfo{author}{\bibfnamefont{A.~H.} \bibnamefont{Mrou{\'e}}},
  \bibinfo{author}{\bibfnamefont{O.}~\bibnamefont{Sarbach}},
  \bibinfo{author}{\bibfnamefont{E.}~\bibnamefont{Schnetter}},
  \bibinfo{author}{\bibfnamefont{M.}~\bibnamefont{Tiglio}}, \bibnamefont{and}
  \bibinfo{author}{\bibfnamefont{M.}~\bibnamefont{Wagman}},
  \bibinfo{journal}{Physical Review D} \textbf{\bibinfo{volume}{85}},
  \bibinfo{pages}{084004} (\bibinfo{year}{2012}).

\bibitem[{\citenamefont{Lindblom et~al.}(2006)\citenamefont{Lindblom, Scheel,
  Kidder, Owen, and Rinne}}]{lial06}
\bibinfo{author}{\bibfnamefont{L.}~\bibnamefont{Lindblom}},
  \bibinfo{author}{\bibfnamefont{M.~A.} \bibnamefont{Scheel}},
  \bibinfo{author}{\bibfnamefont{L.~E.} \bibnamefont{Kidder}},
  \bibinfo{author}{\bibfnamefont{R.}~\bibnamefont{Owen}}, \bibnamefont{and}
  \bibinfo{author}{\bibfnamefont{O.}~\bibnamefont{Rinne}},
  \bibinfo{journal}{Classical and Quantum Gravity}
  \textbf{\bibinfo{volume}{23}}, \bibinfo{pages}{S447} (\bibinfo{year}{2006}).

\bibitem[{\citenamefont{Bernuzzi and Hilditch}(2010)}]{Z4ca}
\bibinfo{author}{\bibfnamefont{S.}~\bibnamefont{Bernuzzi}} \bibnamefont{and}
  \bibinfo{author}{\bibfnamefont{D.}~\bibnamefont{Hilditch}},
  \bibinfo{journal}{Physical Review D} \textbf{\bibinfo{volume}{81}},
  \bibinfo{pages}{084003} (\bibinfo{year}{2010}).

\bibitem[{\citenamefont{Alic et~al.}(2012)\citenamefont{Alic, Bona-Casas, Bona,
  Rezzolla, and Palenzuela}}]{CCZ4a}
\bibinfo{author}{\bibfnamefont{D.}~\bibnamefont{Alic}},
  \bibinfo{author}{\bibfnamefont{C.}~\bibnamefont{Bona-Casas}},
  \bibinfo{author}{\bibfnamefont{C.}~\bibnamefont{Bona}},
  \bibinfo{author}{\bibfnamefont{L.}~\bibnamefont{Rezzolla}}, \bibnamefont{and}
  \bibinfo{author}{\bibfnamefont{C.}~\bibnamefont{Palenzuela}},
  \bibinfo{journal}{Physical Review D} \textbf{\bibinfo{volume}{85}},
  \bibinfo{pages}{064040} (\bibinfo{year}{2012}).

\bibitem[{\citenamefont{Alic et~al.}(2013)\citenamefont{Alic, Kastaun, and
  Rezzolla}}]{CCZ4b}
\bibinfo{author}{\bibfnamefont{D.}~\bibnamefont{Alic}},
  \bibinfo{author}{\bibfnamefont{W.}~\bibnamefont{Kastaun}}, \bibnamefont{and}
  \bibinfo{author}{\bibfnamefont{L.}~\bibnamefont{Rezzolla}},
  \bibinfo{journal}{Phys. Rev. D} \textbf{\bibinfo{volume}{88}},
  \bibinfo{pages}{064049} (\bibinfo{year}{2013}),
  \urlprefix\url{http://link.aps.org/doi/10.1103/PhysRevD.88.064049}.

\bibitem[{\citenamefont{Weyhausen et~al.}(2012)\citenamefont{Weyhausen,
  Bernuzzi, and Hilditch}}]{Z4cb}
\bibinfo{author}{\bibfnamefont{A.}~\bibnamefont{Weyhausen}},
  \bibinfo{author}{\bibfnamefont{S.}~\bibnamefont{Bernuzzi}}, \bibnamefont{and}
  \bibinfo{author}{\bibfnamefont{D.}~\bibnamefont{Hilditch}},
  \bibinfo{journal}{Physical Review D} \textbf{\bibinfo{volume}{85}},
  \bibinfo{pages}{024038} (\bibinfo{year}{2012}).

\bibitem[{\citenamefont{Hilditch et~al.}(2013)\citenamefont{Hilditch, Bernuzzi,
  Thierfelder, Cao, Tichy, and Br{\"u}gmann}}]{Z4cc}
\bibinfo{author}{\bibfnamefont{D.}~\bibnamefont{Hilditch}},
  \bibinfo{author}{\bibfnamefont{S.}~\bibnamefont{Bernuzzi}},
  \bibinfo{author}{\bibfnamefont{M.}~\bibnamefont{Thierfelder}},
  \bibinfo{author}{\bibfnamefont{Z.}~\bibnamefont{Cao}},
  \bibinfo{author}{\bibfnamefont{W.}~\bibnamefont{Tichy}}, \bibnamefont{and}
  \bibinfo{author}{\bibfnamefont{B.}~\bibnamefont{Br{\"u}gmann}},
  \bibinfo{journal}{Physical Review D} \textbf{\bibinfo{volume}{88}},
  \bibinfo{pages}{084057} (\bibinfo{year}{2013}).

\bibitem[{\citenamefont{Laguna and Shoemaker}(2002)}]{LaPS02}
\bibinfo{author}{\bibfnamefont{P.}~\bibnamefont{Laguna}} \bibnamefont{and}
  \bibinfo{author}{\bibfnamefont{D.}~\bibnamefont{Shoemaker}},
  \bibinfo{journal}{Classical and Quantum Gravity}
  \textbf{\bibinfo{volume}{19}}, \bibinfo{pages}{3679} (\bibinfo{year}{2002}).

\bibitem[{\citenamefont{Cao et~al.}(2008)\citenamefont{Cao, Yo, and
  Yu}}]{czyy08}
\bibinfo{author}{\bibfnamefont{Z.}~\bibnamefont{Cao}},
  \bibinfo{author}{\bibfnamefont{H.-J.} \bibnamefont{Yo}}, \bibnamefont{and}
  \bibinfo{author}{\bibfnamefont{J.-P.} \bibnamefont{Yu}},
  \bibinfo{journal}{Physical Review D} \textbf{\bibinfo{volume}{78}},
  \bibinfo{pages}{124011} (\bibinfo{year}{2008}).

\bibitem[{\citenamefont{Etienne et~al.}(2014)\citenamefont{Etienne, Baker,
  Paschalidis, Kelly, and Shapiro}}]{etienne2014improved}
\bibinfo{author}{\bibfnamefont{Z.~B.} \bibnamefont{Etienne}},
  \bibinfo{author}{\bibfnamefont{J.~G.} \bibnamefont{Baker}},
  \bibinfo{author}{\bibfnamefont{V.}~\bibnamefont{Paschalidis}},
  \bibinfo{author}{\bibfnamefont{B.~J.} \bibnamefont{Kelly}}, \bibnamefont{and}
  \bibinfo{author}{\bibfnamefont{S.~L.} \bibnamefont{Shapiro}},
  \bibinfo{journal}{arXiv preprint arXiv:1404.6523}  (\bibinfo{year}{2014}).

\bibitem[{\citenamefont{Marronetti et~al.}(2008)\citenamefont{Marronetti,
  Tichy, Bruegmann, Gonzalez, and Sperhake}}]{MarTicBru08}
\bibinfo{author}{\bibfnamefont{P.}~\bibnamefont{Marronetti}},
  \bibinfo{author}{\bibfnamefont{W.}~\bibnamefont{Tichy}},
  \bibinfo{author}{\bibfnamefont{B.}~\bibnamefont{Bruegmann}},
  \bibinfo{author}{\bibfnamefont{J.}~\bibnamefont{Gonzalez}}, \bibnamefont{and}
  \bibinfo{author}{\bibfnamefont{U.}~\bibnamefont{Sperhake}},
  \bibinfo{journal}{Physical Review D} \textbf{\bibinfo{volume}{77}},
  \bibinfo{pages}{064010} (\bibinfo{year}{2008}).

\bibitem[{\citenamefont{Hemberger et~al.}(2013)\citenamefont{Hemberger,
  Lovelace, Loredo, Kidder, Scheel, Szil\'agyi, Taylor, and
  Teukolsky}}]{HemLovLor13}
\bibinfo{author}{\bibfnamefont{D.~A.} \bibnamefont{Hemberger}},
  \bibinfo{author}{\bibfnamefont{G.}~\bibnamefont{Lovelace}},
  \bibinfo{author}{\bibfnamefont{T.~J.} \bibnamefont{Loredo}},
  \bibinfo{author}{\bibfnamefont{L.~E.} \bibnamefont{Kidder}},
  \bibinfo{author}{\bibfnamefont{M.~A.} \bibnamefont{Scheel}},
  \bibinfo{author}{\bibfnamefont{B.}~\bibnamefont{Szil\'agyi}},
  \bibinfo{author}{\bibfnamefont{N.~W.} \bibnamefont{Taylor}},
  \bibnamefont{and} \bibinfo{author}{\bibfnamefont{S.~A.}
  \bibnamefont{Teukolsky}}, \bibinfo{journal}{Phys. Rev. D}
  \textbf{\bibinfo{volume}{88}}, \bibinfo{pages}{064014}
  (\bibinfo{year}{2013}),
  \urlprefix\url{http://link.aps.org/doi/10.1103/PhysRevD.88.064014}.

\bibitem[{\citenamefont{Galaviz et~al.}(2010)\citenamefont{Galaviz, Br\"ugmann,
  and Cao}}]{cao10}
\bibinfo{author}{\bibfnamefont{P.}~\bibnamefont{Galaviz}},
  \bibinfo{author}{\bibfnamefont{B.}~\bibnamefont{Br\"ugmann}},
  \bibnamefont{and} \bibinfo{author}{\bibfnamefont{Z.}~\bibnamefont{Cao}},
  \bibinfo{journal}{Phys. Rev. D} \textbf{\bibinfo{volume}{82}},
  \bibinfo{pages}{024005} (\bibinfo{year}{2010}),
  \urlprefix\url{http://link.aps.org/doi/10.1103/PhysRevD.82.024005}.

\bibitem[{\citenamefont{Dain et~al.}(2008)\citenamefont{Dain, Lousto, and
  Zlochower}}]{dain2008extra}
\bibinfo{author}{\bibfnamefont{S.}~\bibnamefont{Dain}},
  \bibinfo{author}{\bibfnamefont{C.~O.} \bibnamefont{Lousto}},
  \bibnamefont{and}
  \bibinfo{author}{\bibfnamefont{Y.}~\bibnamefont{Zlochower}},
  \bibinfo{journal}{Physical Review D} \textbf{\bibinfo{volume}{78}},
  \bibinfo{pages}{024039} (\bibinfo{year}{2008}).

\end{thebibliography}
\end{document}